\documentclass[a4paper,11pt]{article}
\usepackage{lmodern}
\usepackage[utf8]{inputenc}
\usepackage[T1]{fontenc}
\usepackage{amsmath}
\usepackage{amssymb}
\usepackage{mathtools}
\usepackage[english]{babel}
\usepackage{url}
\usepackage{booktabs}
\usepackage{longtable}
\usepackage{cellspace}
\usepackage{makecell}
\usepackage{graphicx}
\usepackage{needspace}
\usepackage{xspace}
\usepackage{setspace}
\usepackage{textcase}
\usepackage{multirow}
\usepackage{subcaption}
\usepackage{jheppub}
\usepackage{setspace}
\usepackage{bm}
\usepackage{scrextend}
\usepackage{tikz}
\usepackage[format=plain,labelfont=bf,textfont=it]{caption}

\usepackage{enumitem}
\setlist[description]{%
	itemsep=0pt,               % space between items
	font={\normalfont\scshape}, % set the label font
}

\usepackage{tikz}
\usetikzlibrary{intersections}
\usetikzlibrary{decorations.markings}
\usetikzlibrary{arrows.meta}
\usetikzlibrary{calc}
\usetikzlibrary{graphs}

\makeatletter
\global\let\tikz@ensure@dollar@catcode=\relax
\makeatother

\usepackage{natbib}

\newcommand{\mycomment}[1]{}

\let\originalleft\left
\let\originalright\right
\renewcommand{\left}{\mathopen{}\mathclose\bgroup\originalleft}
\renewcommand{\right}{\aftergroup\egroup\originalright}

\makeatletter
\providecommand*{\shuffle}{%
	\mathbin{\mathpalette\myshuffle@{}}%
}
\newcommand*{\myshuffle@}[2]{%
	\sbox0{$#1\vcenter{}$}%
	\kern .15\ht0 % side bearing
	\rlap{\vrule height .25\ht0 depth 0pt width 2.5\ht0}%
	\raise.1\ht0\hbox to 2.5\ht0{%
		\vrule height 1.75\ht0 depth -.1\ht0 width .17\ht0 %
		\hfill
		\vrule height 1.75\ht0 depth -.1\ht0 width .17\ht0 %
		\hfill
		\vrule height 1.75\ht0 depth -.1\ht0 width .17\ht0 %
	}%
	\kern .15\ht0 % side bearing
}
\makeatother

\DeclareFontFamily{U}{mathx}{\hyphenchar\font45}
\DeclareFontShape{U}{mathx}{m}{n}{
	<5> <6> <7> <8> <9> <10>
	<10.95> <12> <14.4> <17.28> <20.74> <24.88>
	mathx10
}{}
\DeclareSymbolFont{mathx}{U}{mathx}{m}{n}
\DeclareFontSubstitution{U}{mathx}{m}{n}
\DeclareMathAccent{\widecheck}{0}{mathx}{"71}

\renewcommand{\Re}{\operatorname{Re}}
\renewcommand{\Im}{\operatorname{Im}}

\newcommand{\dd}{\!\mathrm{d}}

\newcommand{\RR}{\mathbb{R}}

\newcommand{\CC}{\mathbb{C}}
\newcommand{\QQ}{\mathbb{Q}}
\newcommand{\ZZ}{\mathbb{Z}}

\newcommand{\SL}{\mathrm{SL}}

\newcommand{\HH}{\mathbb{H}}

\newcommand{\EE}{\mathrm{E}}
\newcommand{\GG}{\mathrm{G}}

\DeclareMathOperator{\ad}{ad}

\newcommand{\smatrix}[1]{\begin{smallmatrix}#1\end{smallmatrix}}
\newcommand{\sbmatrix}[1]{\left[\begin{smallmatrix}#1\end{smallmatrix}\right]}

\newcommand{\ompm}[3]{\omega_{\pm}\! \left[\begin{smallmatrix}#1\\#2\end{smallmatrix};#3\right]}
\newcommand{\omplus}[3]{\omega_{+}\! \left[\begin{smallmatrix}#1\\#2\end{smallmatrix};#3\right]}
\newcommand{\omminus}[3]{\omega_{-}\! \left[\begin{smallmatrix}#1\\#2\end{smallmatrix};#3\right]}

\newcommand{\betaeqv}[1]{
	\beta^\mathrm{eqv}\! \big[\begin{smallmatrix}#1\end{smallmatrix}\big]}

\newcommand{\betaeqvtau}[1]{
	\beta^\mathrm{eqv}\! \big[\begin{smallmatrix}#1\end{smallmatrix}; \tau\big]}
\newcommand{\betaeqvx}[2]{
	\beta^\mathrm{eqv}\! \big[\begin{smallmatrix}#1\end{smallmatrix}; #2\big]}

\newcommand{\betaplus}[1]{
	\beta_\mathrm{+}\! \big[\begin{smallmatrix}#1\end{smallmatrix}\big]}
\newcommand{\betaplustau}[1]{
	\beta_\mathrm{+}\! \big[\begin{smallmatrix}#1\end{smallmatrix}; \tau\big]}

\newcommand{\betaminus}[1]{
	\beta_\mathrm{-}\! \big[\begin{smallmatrix}#1\end{smallmatrix}\big]}
\newcommand{\betaminustau}[1]{
	\beta_\mathrm{-}\! \big[\begin{smallmatrix}#1\end{smallmatrix}; \tau\big]}

\newcommand{\bsvBR}[3]{
	\beta^\mathrm{sv} \! \big[\begin{smallmatrix}#1\\#2\end{smallmatrix};#3\big]}

\newcommand{\cno}[2]{
	{\cal C}\! \left[\begin{smallmatrix}#1\\#2\end{smallmatrix}\right]}

\newcommand{\ddsv}[3]{d^{\rm sv}\! \left[\begin{smallmatrix}#1\\#2\end{smallmatrix};#3\right]}

\newcommand{\ddsvpure}{d^{\rm sv}}

\newcommand{\ccsv}[2]{c^{\rm sv}\! \left[\begin{smallmatrix}#1\\#2\end{smallmatrix}\right]}

\newcommand{\imtaupi}{\bigg(\!\frac{\tau_{2}}{\pi}\!\parbox{4pt}{$\bigg)$}}
\newcommand{\cform}[1]{\operatorname{\mathcal{C}\hspace{-3pt}}\big[
	\protect\begin{smallmatrix}#1\protect\end{smallmatrix}\parbox{4pt}{$\big]$}}
\newcommand{\cforml}[1]{\operatorname{\mathcal{C}\hspace{-3pt}}\Big[
	\protect\begin{smallmatrix}#1\protect\end{smallmatrix}\parbox{4pt}{$\Big]$}}
\newcommand{\cformtri}[3]{\operatorname{\mathcal{C}\hspace{-3pt}}\big[\!
	{\setlength{\tabcolsep}{1pt}\setlength\arrayrulewidth{0.8pt}
		\begin{tabular}{c|c|c}
			$\smatrix{#1}$ & $\smatrix{#2}$ & $\smatrix{#3}$
		\end{tabular}
	}
	\!\parbox{4pt}{$\big]$}}

\newcommand{\cformtril}[3]{\operatorname{\mathcal{C}\hspace{-3pt}}\Big[\!
	{\setlength{\tabcolsep}{1pt}\setlength\arrayrulewidth{0.8pt}
		\begin{tabular}{c|c|c}
			$\smatrix{#1}$ & $\smatrix{#2}$ & $\smatrix{#3}$
		\end{tabular}
	}
	\!\parbox{4pt}{$\Big]$}}

\newcommand{\cformboxl}[4]{\operatorname{\mathcal{C}\hspace{-3pt}}\Big[\!
	{\setlength{\tabcolsep}{1pt}\setlength\arrayrulewidth{0.8pt}
		\begin{tabular}{c|c|c|c}
			$\smatrix{#1}$ & $\smatrix{#2}$ & $\smatrix{#3}$ & $\smatrix{#4}$
		\end{tabular}
	}
	\!\parbox{4pt}{$\Big]$}}

\newcommand{\cformkitel}[5]{\operatorname{\mathcal{C}\hspace{-3pt}}\Big[\!
	{\setlength{\tabcolsep}{1pt}\setlength\arrayrulewidth{0.8pt}
		\begin{tabular}{c|c||c|c||c}
			$\smatrix{#1}$ & $\smatrix{#2}$ & $\smatrix{#3}$ & $\smatrix{#4}$
			& $\smatrix{#5}$
		\end{tabular}
	}
	\!\parbox{4pt}{$\Big]$}}

\newdimen\aboverulesepbuffer
\newdimen\belowrulesepbuffer

\newcommand{\cformtet}[6]{
	\operatorname{\mathcal{C}\hspace{-3pt}}\Bigg[\!
	{\setlength{\aboverulesep}{-1pt}
		\setlength{\belowrulesep}{0pt}
		\setlength{\tabcolsep}{2pt}
		\setlength\arrayrulewidth{0.8pt}
		\begin{tabular}{c||c||c}
			$\begin{smallmatrix}#1\end{smallmatrix}$ &
			$\begin{smallmatrix}#2\end{smallmatrix}$ &
			$\begin{smallmatrix}#3\end{smallmatrix}$ \\[0.7ex]
			\cmidrule(l{2pt}r{5pt}){1-1}\cmidrule(l{2pt}r{5pt}){2-2}
			\cmidrule(l{2pt}r{2pt}){3-3}
			\rule[1.5ex]{0pt}{1ex}$\begin{smallmatrix}#4\end{smallmatrix}$ &
			$\begin{smallmatrix}#5\end{smallmatrix}$ &
			$\begin{smallmatrix}#6\end{smallmatrix}$
	\end{tabular}}
	\!\parbox{4pt}{$\Bigg]$}}

\newcommand{\cformp}[1]{\operatorname{\mathcal{C}^+\hspace{-3pt}}\big[
	\protect\begin{smallmatrix}#1\protect\end{smallmatrix}\parbox{4pt}{$\big]$}}
\newcommand{\cformtrip}[3]{\operatorname{\mathcal{C}^+\hspace{-3pt}}\big[\!
	{\setlength{\tabcolsep}{1pt}\setlength\arrayrulewidth{0.8pt}
		\begin{tabular}{c|c|c}
			$\smatrix{#1}$ & $\smatrix{#2}$ & $\smatrix{#3}$
		\end{tabular}
	}
	\!\parbox{4pt}{$\big]$}}
\newcommand{\cformboxp}[4]{\operatorname{\mathcal{C}^+\hspace{-3pt}}\big[\!
	{\setlength{\tabcolsep}{1pt}\setlength\arrayrulewidth{0.8pt}
		\begin{tabular}{c|c|c|c}
			$\smatrix{#1}$ & $\smatrix{#2}$ & $\smatrix{#3}$ & $\smatrix{#4}$
		\end{tabular}
	}
	\!\parbox{4pt}{$\big]$}}

\newcommand{\cformkitep}[5]{\operatorname{\mathcal{C}^+\hspace{-3pt}}\big[\!
	{\setlength{\tabcolsep}{1pt}\setlength\arrayrulewidth{0.8pt}
		\begin{tabular}{c|c||c|c||c}
			$\smatrix{#1}$ & $\smatrix{#2}$ & $\smatrix{#3}$ & $\smatrix{#4}$
			& $\smatrix{#5}$
		\end{tabular}
	}
	\!\parbox{4pt}{$\big]$}}
\newcommand{\cformtetp}[6]{
	\operatorname{\mathcal{C}^+\hspace{-3pt}}\Bigg[\!
	{\setlength{\aboverulesep}{-1pt}
		\setlength{\belowrulesep}{0pt}
		\setlength{\tabcolsep}{2pt}
		\setlength\arrayrulewidth{0.8pt}
		\begin{tabular}{c||c||c}
			$\begin{smallmatrix}#1\end{smallmatrix}$ &
			$\begin{smallmatrix}#2\end{smallmatrix}$ &
			$\begin{smallmatrix}#3\end{smallmatrix}$ \\[0.7ex]
			\cmidrule(l{2pt}r{5pt}){1-1}\cmidrule(l{2pt}r{5pt}){2-2}
			\cmidrule(l{2pt}r{2pt}){3-3}
			\rule[1.5ex]{0pt}{1ex}$\begin{smallmatrix}#4\end{smallmatrix}$ &
			$\begin{smallmatrix}#5\end{smallmatrix}$ &
			$\begin{smallmatrix}#6\end{smallmatrix}$
	\end{tabular}}
	\!\parbox{4pt}{$\Bigg]$}}

\tikzset{
	gluon/.style={decoration={coil,
			pre length=2pt,post length=2pt,segment length=3pt},decorate},
	vertex/.style={circle,inner sep=1.5},
	vertexdot/.style={circle,draw,fill,inner sep=0.7},
	label/.style={fill=white,font=\footnotesize},
	mlabel/.style={outer sep=2,font=\footnotesize},
	biglabel/.style={fill=white,font=\normalsize},
	tlabel/.style={fill=white,font=\footnotesize,inner sep=0pt},
	edge/.style={line width=0.5},
	dedge/.style={line width=0.5,decoration=
		{markings,mark=at position #1 with {\arrow[scale=1.3]{latex}}},
		postaction=decorate},
	dedge/.default=0.8,
	mgf/.style={baseline=-0.6ex}}

\newcommand{\nwc}{\newcommand}
\nwc{\ba}  {\begin{array}}
	\nwc{\ea}  {\end{array}}
\nwc{\bdm} {\begin{displaymath}}
	\nwc{\edm} {\end{displaymath}}
\nwc{\bda} {\bdm\ba{lcl}} 
\nwc{\eda} {\ea\edm}
\nwc{\bc}  {\begin{center}}
	\nwc{\ec}  {\end{center}}
\nwc{\ds}  {\displaystyle}
\nwc{\bmat}{\left(\ba}
\nwc{\emat}{\ea\right)}
\nwc{\nnn} {\nonumber \vspace{.2cm} \\ }
\nwc{\ra}  {\rightarrow}
\nwc{\lra} {\longrightarrow}
\nwc{\p} {\partial}
\nwc{\rcr} {\nabla_{\rm alt}}
\nwc{\barrcr} {\overline{\nabla_{\rm alt}}}
\nwc{\ep} {\epsilon}

\def\ad{\mathrm{ad}}
\def\dd{\text{d}}

\def\ad{\mathrm{ad}}
\def\dd{\text{d}}

\newcommand{\zetaT}{z}
\newcommand{\sigmaT}{\sigma}

\title{\boldmath From Modular Graph Forms to Iterated Integrals}

\author[a]{E. Claasen}
\author[b]{M. Doroudiani}
\affiliation[a]{Max-Planck-Institut f\"ur Gravitationsphysik (Albert-Einstein-Institut), Am M\"uhlenberg 1, DE-14476 Potsdam, Germany}
\affiliation[b]{School of Physics \& Astronomy, University of Southampton, SO17 1BJ, UK}

\emailAdd{emiel.claasen@aei.mpg.de}
\emailAdd{m.doroudiani@soton.ac.uk}

\abstract{Modular graph forms are a class of non-holomorphic modular forms that arise in the low-energy expansion of genus-one closed string amplitudes. In this work, we introduce a systematic procedure to convert lattice-sum representations of modular graph forms into iterated integrals of holomorphic Eisenstein series and provide a \textsc{Mathematica} package that implements all modular graph form topologies up to four vertices. To achieve this, we introduce specific tree-representations of modular graph forms. The presented method enables the conversion of the integrand of the four-graviton one-loop amplitude in Type II superstring theory at eighth order in the inverse string tension $\alpha^{\prime 8}$, which we use to calculate the $\alpha^{\prime 8}\zeta_3\zeta_5$ contribution to the analytic part of the amplitude.}
\makeatletter
\gdef\@fpheader{}
\makeatother
\begin{document}
	\maketitle
	\flushbottom
	
	\section{Introduction}
	Modular Graph Forms (MGFs) are a class of non-holomorphic modular forms associated to directed simple\footnote{A simple graph is a graph without edges from a vertex to itself, i.e. without (self-)loops.} graphs with each edge labeled by two integers called the holomorphic and anti-holomorphic edge labels. They originated from the diagrammatic representation of contributions to the low-energy expansion of closed string amplitudes at genus one in Type II superstring theory \cite{Green:2000,Green:2008uj,DHoker:2015gmr,DHoker:2019blr,DHoker:2015wxz} and were subsequently generalized to modular forms associated to any directed labeled simple graph in \cite{DHoker:2016mwo}. Modular graph forms are graphical representations of lattice sums. The study of these lattice sums revealed many identities between different MGFs \cite{DHoker:2015gmr,DHoker:2015sve,DHoker:2016mwo,DHoker:2016quv, Gerken:2018zcy,Basu:2016kli,Gerken:2020aju,Kleinschmidt:2017ege}, which are not directly apparent from their lattice-sum representation. Several tools have been developed to generate these identities, with the most important ones being momentum conservation, holomorphic subgraph reduction, Fay identities and the sieve algorithm \cite{DHoker:2016mwo,Gerken:2018zcy,Gerken:2020aju,DHoker:2015sve,DHoker:2016quv}. 
	
	Infinite families of non-holomorphic modular forms that arise from iterated integrals of holomorphic Eisenstein series and their complex conjugates were constructed in \cite{Brown:2017qwo,Brown:2017qwo2,Broedel:2018izr,Gerken:2019cxz,Gerken:2020yii,drewitt2022laplace,Dorigoni:2022npe,Dorigoni:2024oft}, where the number of integrated holomorphic Eisenstein series is called the depth of the iterated integral and the sum of the modular weights of the holomorphic Eisenstein series is called its degree. In \cite{Brown:2017qwo} it was found that a specific combination of iterated Eisenstein integrals obeyed the same Laplace equation as a certain MGF, showing their equivalence up to a constant. This suggested that these iterated Eisenstein integrals contain MGFs. The dictionary between iterated Eisenstein integrals and MGFs was expanded in \cite{Broedel:2018izr} using elliptic multiple zeta values. Later, by using the generating series of MGFs, a systematic expansion of dihedral MGFs in terms of the iterated integrals was found \cite{Gerken:2020yii}. The bottleneck that prevents an extension of this method to MGFs with more vertices is the need of the degeneration limit $\tau\rightarrow i\infty$ of the generating series, which is only known for the dihedral case \cite{Gerken:2020yii}. Instead, using the recently developed modular version of iterated Eisenstein integrals and their differential equations \cite{ Brown:2017qwo,Brown:2017qwo2,Gerken:2020yii,drewitt2022laplace,Dorigoni:2022npe,Dorigoni:2024oft}, we present an algorithmic procedure to expand any MGF free of holomorphic and antiholomorphic subgraphs in terms of equivariant iterated Eisenstein integrals.
	
	The main idea of this algorithm is based on the sieve algorithm \cite{DHoker:2016mwo} and its modifications in \cite{Gerken:2020xte,Hidding:2022vjf}. It roughly works as follows. Given an MGF, take its holomorphic derivative and afterwards use the identities mentioned before to get rid of problematic terms that would obstruct further application of the algorithm. In certain cases, this can produce new MGFs that factorize into a holomorphic Eisenstein series and another MGF. Instead of canceling the terms involving holomorphic Eisenstein series using derivatives of other MGFs as one does in the original sieve algorithm \cite{DHoker:2016mwo}, remove these holomorphic Eisenstein series manually and act again with the holomorphic derivative on all the remaining MGFs \cite{Gerken:2020xte} (including the coefficients of the holomorphic Eisenstein series). Getting rid of problematic terms and removing the holomorphic Eisenstein series again, we can iterate this procedure until no MGFs besides holomorphic Eisenstein series remain. We can then integrate this system back up to get an iterated integral description of the original MGF. A similar algorithm was used to convert elliptic MGFs into iterated integrals in \cite{Hidding:2022vjf}. 
	
	In this paper, we demonstrate how this algorithm can be realized explicitly. The differentiation part of the algorithm can conveniently be represented by a tree, where for every removed holomorphic Eisenstein series, we introduce a new branch of the tree and leave the holomorphic Eisenstein series at this new branch. We can now undo all the derivatives we performed by integrating back up. This tree provides us with the instructions on how to perform the integration procedure, with the holomorphic Eisenstein series providing the integration kernels. The integration procedure can be reduced to a linear algebra problem due to the differential equation that the iterated integrals satisfy. This gives a representation of MGFs in terms of iterated integrals of holomorphic Eisenstein series. These tree-representations resemble the way symbols are constructed of QFT amplitudes \cite{Goncharov:2010jf}.
	
	Naively, one may think that too much information is lost after taking derivatives, such that integrating back up gives a rather incomplete representation due to the large amount of integration constants. However, using the language of equivariant iterated integrals to integrate back up fixes most of them because of their modular properties. The only case when this fails is whenever we encounter modular invariant integrals \cite{Brown:2017qwo,DHoker:2016mwo}. Adding a constant to a modular invariant expression leaves it modular invariant, which means that two related definitions of modular invariant objects might differ by a constant. In this case, we have to match the constants in the cuspidal expansions of the MGF with the iterated integral representation to fix the integration constants. These constants are rational linear combinations of single-valued Multiple Zeta Values (svMZVs) \cite{DHoker:2015wxz,Dorigoni:2024oft,Zerbini:2015rss}. The specific svMZVs that arise can be inferred from uniform transcendentality of the cuspidal expansions \cite{Zerbini:2015rss,DHoker:2015wxz}, which leaves only their rational coefficients undetermined. Calculation of these cuspidal expansions turns out to be the only real bottleneck of the algorithm.
	
	\subsection{Discussion and outlook}
	The algorithm presented in this paper is in principle applicable to any convergent modular graph form free of holomorphic and antiholomorphic subgraphs and has been implemented in the \textsc{Mathematica} package \textsc{MGFtoBeqv}. This package resides in a git repository at \url{https://github.com/emieltcc/MGFtoBeqv}, which comes with a user manual. However, there are two obstacles that limit the applicability of the algorithm to any MGF. 
	
	One of them is the lack of available data to fix specific integration constants in the algorithm. As discussed in the introduction, in order to turn the tree-representation into an iterated integral, we have to perform multiple integrals to undo the differential operations. All integration constants are fixed by modularity, except whenever the modular weight of an integral vanishes. In those specific instances, we need the data of the constants in the Laurent polynomial for both the lattice-sum and the iterated Eisenstein integral representations. These are known for lattice-sum representations of MGFs for $|A|+|B|\leq 12$ and only in some instances for $|A|+|B|>12$, with $|A|$ and $|B|$ the sums of all holomorphic and antiholomorphic edge labels respectively\footnote{The sum of the edge labels $|A|+|B|$ in the lattice-sum representation of a particular MGF equals the degrees of the iterated integrals involved in the iterated Eisenstein integral representation of that MGF.}, all for $n$-vertex graphs with $n\leq 3$ \cite{Zerbini:2015rss,Green:2008uj,Gerken:2020aju,DHoker:2019xef,Zagier:2019eus,Vanhove:2020qtt,Gerken:2020yii}. Similarly, they are known for the equivariant iterated integrals of all degrees at depth one and two, and degrees $\leq 20$ at depth 3 \cite{Dorigoni:2021jfr,Dorigoni:2021ngn,Broedel:2018izr,Dorigoni:2024iyt,Dorigoni:2024oft}, and can be computed from appendix \ref{appendix: LP}. For depth 4 at degree 16, the constants proportional to $\zeta_3\zeta_5$ are calculated in appendix \ref{appendix: csv}. Because of these limitations, we decided to limit the applicability of the package to MGFs with anti-holomorphic weight up to $|B|=11$. Nonetheless, even without this data, it is possible to perform the algorithm and leave the rational coefficients of the svMZVs undetermined in the final expression. 
	
	The other limitation is the lack of implemented identities between MGFs in the \textsc{Mathematica} package of \cite{Gerken:2020aju} for graphs with four or more vertices. However, other than the lack of data for the constants in the Laurent polynomials, this obstruction is purely an inconvenience. There is no reason for higher-vertex MGFs to not be convertible to iterated integral descriptions, but the computations become extremely laborious without these computer-automated simplifications. Therefore, we have implemented specific identities for graphs with up to four vertices in the \textsc{MGFtoBeqv} package to facilitate the conversion of four-vertex graphs. This allowed us to convert all modular graph functions that appear in the one-loop four-graviton amplitude in type II superstring theory up to seventh order in the inverse string tension $\alpha^\prime$ in the low-energy expansion to iterated integrals and compute their integral over the fundamental domain and hence the amplitude, which was done in \cite{Claasen:2024ssh}. The iterated integral representations of those modular graph functions can be found in the ancillary file of \cite{Claasen:2024ssh}. Furthermore, in this paper we give the full integrand at eighth order in $\alpha^\prime$ by converting all MGFs to iterated integrals where the combination of missing constants was fixed by a correspondence with UV-divergences in effective field theory one-loop matrix elements \cite{Edison:2021ebi}.
	
	If MGF identities (specifically momentum conservation and holomorphic subgraph reduction) for higher-vertex topologies would be implemented in the future, the algorithm could be used to convert the integrand of higher-point scattering amplitudes, e.g. the five-graviton superstring amplitude to iterated Eisenstein integrals. The time it takes the algorithm to convert an MGF with $|A|+|B|\leq 16$ on a local computer is of the order of split-seconds to a few minutes. For example, converting all four-graviton modular graph functions that contribute to the $\mathcal{D}^{14}\mathcal{R}^4$ term in the effective action of Type II string theory took less than 10 minutes on a standard laptop. Therefore, computation time would not be an issue for higher-point amplitudes for relatively low weights.
	
	Another interesting direction for future research would be to classify all modular graph forms up to depth 3 and to provide general formulas for their integration over the fundamental domain, which is under control up to depth 3. The tree-representations can be used to diagnose the drop in complexity from loop order to iterated integral depth. It would be interesting to see if there are any general patterns with respect to the correspondence between loop order, topology and depth.
	
	\subsection{Outline}
	This work is structured as follows. In section \ref{section: MGFs} we review lattice-sum representations of modular graph forms and techniques to produce identities between MGFs, which we use to construct trees in section \ref{subsection: trees}. In section \ref{section: Iterint}, we review iterated integrals of holomorphic Eisenstein series and their modular completions. These satisfy a differential equation that will be introduced in \ref{subsection: diffeqn}, which plays an important role in the algorithm. In section \ref{section: alg}, we explain how to translate lattice sums into iterated integrals using the tools introduced in sections \ref{section: MGFs}, \ref{subsection: trees} and \ref{section: Iterint}. As an application, we write the integrand of the one-loop four-graviton superstring amplitude at order $\alpha^{\prime 8}$ in terms of iterated integrals in section \ref{subsection: w8}. In appendix \ref{appendix: LP} we give a general formula for the Laurent polynomial of equivariant iterated Eisenstein integrals, which we apply to specific depth 4 examples relevant to integrand at order $\alpha^{\prime 8}$ in appendix \ref{appendix: csv}.

	\section{Review of modular graph forms and functions}
	\label{section: MGFs}
	Modular Graph Forms (MGFs) originated from integrals over marked points on a torus \cite{Green:2000,Green:2008uj,DHoker:2015gmr,DHoker:2019blr,DHoker:2015wxz}, but their definition can be extended to any labeled directed simple graph \cite{DHoker:2016mwo}. We can associate to such a graph a lattice sum that transforms as a modular form, which is called a modular graph form. We begin this section by introducing MGFs in section \ref{subsection: mgfintro}. There exists a wealth of identities and differential relations between MGFs \cite{DHoker:2015gmr,DHoker:2015sve,DHoker:2016mwo,DHoker:2016quv, Gerken:2018zcy,Basu:2016kli,Gerken:2020aju,Kleinschmidt:2017ege}, certain classes of which will be discussed in section \ref{subsec: identities}. MGFs have been reviewed in several places in the literature including \cite{DHoker:2022dxx}, we mainly follow \cite{Gerken:2020xte}.
	
	\subsection{Lattice-sum representations}
	\label{subsection: mgfintro}
	Modular graph forms form a subset of more general objects called non-holomorphic modular forms that we first introduce. Consider the modular group $\rm{PSL}(2,\ZZ)$ acting on the complex upper half plane $\HH:=\{z\in\CC\mid\Im z>0\}$ via modular transformations
	\begin{equation}
		\tau\mapsto \frac{\alpha\tau +\beta}{\gamma\tau+\delta}\,,\qquad \begin{pmatrix}
			\alpha & \beta \\
			\gamma & \delta 
		\end{pmatrix}\in\rm{PSL}(2,\mathbb{Z})\,,\qquad \tau=\tau_1+\text{$i$} \tau_2\in\HH\,,
		\label{eq: tautransf}
	\end{equation}
	A holomorphic function $f:\mathbb{H}\rightarrow \mathbb{C}$ that transforms under modular transformations as
	\begin{equation}
		f\left(\frac{\alpha\tau+\beta}{\gamma\tau +\delta}\right)=(\gamma\tau+\delta)^af(\tau)\,,
		\label{eq: holmodform}
	\end{equation}
	and satisfies a suitable growth condition as $\tau_2\rightarrow \infty$ \cite{serre2012course}, is called a holomorphic modular form of weight $a$. In particular, due to the transformation property (\ref{eq: holmodform}), it is invariant under $\tau\mapsto\tau+1$. This means we can Fourier expand a holomorphic modular form as
	\begin{align}
		f(\tau)=\sum_{m=0}^\infty a_n q^n\,,
		\label{eq: fourierholmodform}
	\end{align}
	where $q=e^{2\pi i \tau}$ and $a_n$ are $\tau_2$-independent coefficients as a result of holomorphicity. We call a holomorphic modular form with $a_0=0$ a holomorphic cusp form.
	
	Another definition is that of a non-holomorphic modular form. A real-analytic function $f:\HH\rightarrow\CC$ that transforms under modular transformations as
	\begin{align}
		f\left(\frac{\alpha\tau+\beta}{\gamma\tau +\delta}\right)=(\gamma\tau+\delta)^a(\gamma\Bar{\tau}+\delta)^bf(\tau)\,,
		\label{eq: nonholmodform}
	\end{align}
	and that has $q$-expansion
	\begin{align}
		f(\tau)=\sum_{m,n=0}^\infty a_{m,n}(\tau_2)q^m\Bar{q}^n\,,
		\label{eq: fouriernonholmodform}
	\end{align}
	is called a non-holomorphic modular form of modular weights $(a,b)$ and of total modular weight $a+b$. We call the entries $a$ and $b$ the holomorphic and anti-holomorphic weights respectively. Further, in the case of MGFs, the $a_{m,n}(\tau_2)$ are Laurent polynomials in $\tau_2$. The Fourier zero mode of this expansion contains two parts: a Laurent polynomial in $\tau_2$ given by $a_{0,0}$ and exponentially suppressed terms for $m=n\neq 0$\footnote{Another frequently used term for the exponentially suppressed terms is the rapid decay part of the modular form.}. A (non-holomorphic) modular form with vanishing modular weights is called a modular function.
	
	Let us see some illustrative examples of these different types of modular forms. A holomorphic modular form that plays an essential role in the algorithm is the holomorphic Eisenstein series of weight $k\geq 4$ even, defined by
	\begin{align}
		\GG_k(\tau)=\sum_{(m,n)\in\ZZ^2}^\prime\frac{1}{(m\tau+n)^k}=2\zeta_k+\frac{2(2\pi i)^{k-1}}{(k-1)!}\sum_{n>0}\sigma_{k-1}(n)q^n\,,
		\label{eq: holEis}
	\end{align}
	where in the second equality we give its Fourier expansion as in (\ref{eq: fourierholmodform}). Here $\zeta_k\coloneqq \sum_{n>0}n^{-k}$ is the Riemann zeta function evaluated at $k$ and $\sigma_{k-1}(n)\coloneqq\sum_{d|n} d^{k-1}$ is the divisor sum. The primed summation implies the exclusion of $(m,n)=(0,0)$. Note that for $k$ odd, the lattice sum representation of $\GG_k$ clearly vanishes, while its Fourier expansion does not, allowing for a non-vanishing extension of the definition to odd $k$. A very important example of a non-holomorphic modular form is $\tau_2=\Im \tau$, which has weight $(-1,-1)$. Another prominent example is that of the non-holomorphic (real-analytic) Eisenstein series, which is a modular function defined by
	\begin{equation}
		\EE_s(\tau)=\left(\frac{\tau_2}{\pi}\right)^s\sum_{(m,n)\in\mathbb{Z}^2}^\prime\frac{1}{|m\tau+n|^{2s}}\,,
		\label{eq: nonholEis}
	\end{equation}
	with $s\in\mathbb{C}$ and $\Re(s)>1$ for convergence.
	
	Besides these examples, there exists a class of non-holomorphic modular forms called modular graph forms; non-holomorphic modular forms associated to directed labeled simple graphs. We define these as follows. Consider a directed simple graph $\Gamma$ where each edge $e$ is labeled by a pair of integers $(a_e,b_e)$ called the holomorphic and anti-holomorphic labels of the edge $e$ respectively. We assign to any such graph the following series
	\begin{equation}
		\mathcal{C}_\Gamma(\tau)=\sum_{p_e}^\prime \prod_{e\in E_\Gamma}\frac{1}{p_e^{a_e}\Bar{p}_e^{b_e}}\prod_{i\in V_\Gamma}\delta\left(\sum_{e'\in E_\Gamma}\Gamma_{ie' }p_{e'}\right)\,,
		\label{eq: mgfdef}
	\end{equation}
	where $E_\Gamma$ is the edge set of $\Gamma$, $V_\Gamma$ the vertex set, $\Gamma_{ie}$ is the incidence matrix. The dependence on the modular parameter $\tau$ is through the sum over lattice momenta $p_e$ taking values in the lattice $\Lambda=(\mathbb{Z}+\tau\mathbb{Z})\backslash\{0\}$. Here $\delta$ denotes the Kronecker delta in the following sense $\delta\left(m\tau+n\right) \coloneqq \delta(m)\delta(n)$ with the $\delta$'s on the right-hand side being the usual Kronecker deltas for $m,n\in\mathbb{Z}$. Equation (\ref{eq: mgfdef}) defines a modular form called a modular graph form of modular weights $(|A|,|B|):=(\sum_{e\in E_\Gamma} a_e,\sum_{e\in E_\Gamma} b_e)$ and is called a Modular Graph Function (MGF\footnote{Note that we refer to both modular graph forms and modular graph functions with the same acronym MGF. Whenever we specifically want to mention either of them, we write them out in full.}) if $|A|=|B|=0$. We list some important properties of MGFs:
	\begin{itemize}
		\item $\mathcal{C}_\Gamma(\tau)=0$ for odd total modular weight $|A|+|B|$.
		\item Swapping the direction of an edge $e$ amounts to a sign $(-1)^{a_e+b_e}$ in $\mathcal{C}_\Gamma(\tau)$.
		\item If $\Gamma$ contains a cut-edge\footnote{A cut-edge of a graph $\Gamma$ is an edge that when removed increases the number of connected components of $\Gamma$.}, then $\mathcal{C}_\Gamma(\tau)=0$ due to the momentum-conserving delta function in (\ref{eq: mgfdef}) at the vertices attached to that cut-edge.
		\item Vertices of degree two can be removed by adding the weights of the adjacent edges to form one single edge.
		\item If $\Gamma$ consists of multiple connected components, $\mathcal{C}_\Gamma(\tau)$ equals the product of the MGFs for all components of $\Gamma$. Similarly if $\Gamma$ contains cut-vertices, then $\mathcal{C}_\Gamma(\tau)$ equals the product of the component MGFs (containing the cut-vertices) that become disconnected by removing that vertex.
		\item Complex conjugation of $\mathcal{C}_\Gamma(\tau)$ amounts to swapping all labels $a_e\leftrightarrow b_e$.
	\end{itemize}
	In general, (\ref{eq: mgfdef}) does not converge for arbitrary labels $(a_e,b_e)$. Conditions on convergence have been investigated in \cite{Gerken:2020aju,DHoker:2017zhq} and are functions of these labels only. We comment on divergent graphs in the algorithm at the end of this section and in section \ref{subsec: identities}. 
	
	Using a slight abuse of notation, we sometimes refer to MGFs with the graphs themselves. In this way, we have for example
	\tikzset{->-/.style={decoration={
				markings,
				mark=at position .9 with {\arrow{latex}}},postaction={decorate}}}
	\begin{equation}
		\begin{tikzpicture}[baseline={([yshift=-1ex]current bounding box.center)}]
			\node (1) at (0,0) {$1$};
			\node (2) at (3.5,0) {$2$};
			\draw[->-] (1) to[bend left=50]
			node[label,fill=white]{$(k,0)$} (2);
			\draw[->-] (1) to[bend right=40]
			node[label,fill=white]{$(0,0)$} (2);
		\end{tikzpicture}:=\sum_{p}^\prime \frac{1}{p^k}=\GG_k\,,
		\label{eq: dihedral}
	\end{equation}
	which is the holomorphic Eisenstein series from (\ref{eq: holEis}). Further, it is convenient to classify MGFs according to the number of vertices and to introduce specific notations for them. As monohedral MGFs vanish, the simplest non-vanishing instance is that of a dihedral MGF, for which we introduce the following notation \cite{Gerken:2020aju}
	\begin{equation}
		\cforml{a_1&\dots&a_r\\b_1&\dots&b_r}:= \begin{tikzpicture}[baseline={([yshift=-1ex]current bounding box.center)}]
			\node (1) at (0,0) {$1$};
			\node (2) at (3.5,0) {$2$};
			\draw[->-] (1) to[bend left=50]
			node[label,fill=white]{$(a_{1},b_{1})$} (2);
			\draw[->-] (1) to[bend left=20]
			node[label,fill=white]{$(a_{2},b_{2})$} (2);
			\node at (1.75,-0.1) {$\vdots$};
			\draw[->-] (1) to[bend right=40]
			node[label,fill=white]{$(a_{r},b_{r})$} (2);
		\end{tikzpicture}=\sum_{p_1,\dots,p_r}^\prime\frac{\delta(p_1+\dots+p_r)}{p_1^{a_1}\Bar{p}_1^{b_1}\dots p_r^{a_r}\Bar{p}_r^{b_r}}\,.
		\label{eq: dihedralC}
	\end{equation}
	We can generalize the notation to graphs with more vertices. First we define $\cform{A\\B}:=\cform{a_1&\dots&a_r\\b_1&\dots&b_r}$, i.e. we group parallel edges in blocks with $A,B$ being $r$-dimensional row-vectors. The next class of MGFs consists of trihedral graphs, which we denote as follows
	\begin{equation}
		\cformtril{A_1\\B_1}{A_2\\B_2}{A_3\\B_3}=\begin{tikzpicture}[baseline=(zero.base)]
			\node (1) at (0,0) {$1$};
			\node (2) at (60:3.5) {$2$};
			\node (3) at (3.5,0) {$3$};
			\draw[->-] (1) to[bend left=15] (2);
			\draw[->-] (1) to[bend right=15] (2);
			\draw[->-] (1) to node(zero)[fill=white]{$\sbmatrix{A_{1}\\B_{1}}$}(2);
			\draw[->-] (2) to[bend left=15] (3);
			\draw[->-] (2) to[bend right=15] (3);
			\draw[->-] (2) to node[fill=white]{$\sbmatrix{A_{2}\\B_{2}}$}(3);
			\draw[->-] (3) to[bend left=15] (1);
			\draw[->-] (3) to[bend right=15] (1);
			\draw[->-] (3) to node[fill=white]{$\sbmatrix{A_{3}\\B_{3}}$}(1);
		\end{tikzpicture}\,.
		\label{eq: trihedralC}
	\end{equation}
	For each of the higher-vertex cases, we have several distinct topologies for which we could introduce different notations. We refer to \cite{Gerken:2020aju} for more details on the case of notations for four-vertex MGFs. As dihedral, trihedral and four-vertex graphs already cover all cases relevant to four-graviton string scattering \cite{Green:2000,Green:2008uj,DHoker:2015gmr, DHoker:2019blr,DHoker:2015wxz,Claasen:2024ssh}, higher-vertex notations have not been introduced. Nevertheless, the algorithm itself is applicable to graphs with any amount of vertices.
	
	Having defined modular (graph) forms, we can define covariant differential operators that act on them. These are known as Maa{\ss} or Cauchy-Riemann operators \cite{Maass} defined by
	\begin{align}
		\nabla^{(a)}=2i\tau_2\partial_\tau+a,\qquad \overline{\nabla}^{(b)}=-2i\tau_2\partial_{\Bar{\tau}}+b\,.
		\label{eq: Maass}
	\end{align}
	These act on modular forms of weights $(a,b)$ as 
	\begin{equation}
		\nabla^{(a)}:(a,b)\rightarrow (a+1,b-1)\,,\qquad \overline{\nabla}^{(b)}:(a,b)\rightarrow (a-1,b+1)\,,
		\label{eq: maassact}
	\end{equation}
	i.e. they act covariantly on modular forms and can be interpreted as raising and lowering operators of $\SL(2,\RR)$. To find their explicit actions on modular graph forms, we use 
	\begin{align}
		\nabla^{(a)}\left(\frac{1}{p^a}\right)=\frac{a}{p^{a+1}\Bar{p}^{-1}}\,,\qquad\overline{\nabla}^{(b)}\left(\frac{1}{\bar p^b}\right)=\frac{b}{p^{a-1}\Bar{p}^{b+1}}\,,
		\label{eq: diffaction}
	\end{align}
	such that
	\begin{align}
		\nabla^{(|A|)}\mathcal{C}_\Gamma=\sum_{e\in E_\Gamma}a_e\mathcal{C}_{\Gamma_{(a_e,b_e)\rightarrow (a_e+1,b_e-1)}}\,,\qquad  \overline{\nabla}^{(|B|)}\mathcal{C}_\Gamma=\sum_{e\in E_\Gamma}b_e\mathcal{C}_{\Gamma_{(a_e,b_e)\rightarrow (a_e-1,b_e+1)}}\,.
		\label{eq: nablaonMGF}
	\end{align}
	It is clear from the formulas that the action of such a differential operator does not depend on the topology of a graph. On a single block (dihedral MGF) it acts as 
	\begin{align}
		\nabla^{(|A|)}\cforml{A\\B}=\sum_{e=1}^ra_e\cforml{A+S_e\\B-S_e}\,,\qquad \overline{\nabla}^{(|B|)}\cforml{A\\B}=\sum_{e=1}^rb_e\cforml{A-S_e\\B+S_e}\,,
		\label{eq: nablaondihedral}
	\end{align}
	where the $j^\mathrm{th}$ entry of the row vector $S_e$ is $\delta_{ej}$ \cite{DHoker:2016mwo}. This generalizes using (\ref{eq: nablaonMGF}) to graphs of different topologies. As we can see from (\ref{eq: nablaonMGF}), the action of the Maa{\ss} operators on $\mathcal{C}_\Gamma(\tau)$ change the holomorphic label of a given edge by $\pm 1$ while the corresponding anti-holomorphic label will simultaneously change by $\mp1$. As the convergence properties of MGFs depend only on their sum \cite{Gerken:2020aju}, acting with these operators on convergent MGFs cannot produce divergent MGFs.
	
	\subsection{Identities between MGFs}
	\label{subsec: identities}
	The properties of MGFs listed in the previous section already introduce relations between seemingly different MGFs. It turns out that MGFs satisfy many more algebraic identities that do not follow from any of the previously listed properties and are not directly apparent from their lattice-sum representations \cite{DHoker:2015gmr,DHoker:2015sve,DHoker:2016mwo,DHoker:2016quv, Gerken:2018zcy,Basu:2016kli,Gerken:2020aju,Kleinschmidt:2017ege}. Some simple examples include \cite{Zagierunpub,DHoker:2015gmr,DHoker:2016mwo}
	\begin{equation}    \imtaupi^3\cform{1\,1\,1\\1\,1\,1}=\EE_3+\zeta_3,\qquad \imtaupi^5\cformtri{1\,1\\1\,1}{1\,1\\1\,1}{1\\1}=2\imtaupi^5\cform{3\,1\,1\\3\,1\,1}-\frac{2}{5}\EE_5-\frac{3}{10}\zeta_5\,.
		\label{eq: simpleids}
	\end{equation}
	These type of identities arise from manipulating MGFs using methods of which we will introduce the ones relevant to the conversion algorithm: momentum conservation, Holomorphic Subgraph Reduction (HSR) and Fay identities. Their importance to the algorithm will be briefly mentioned below and discussed in more detail in section \ref{subsection: trees}.
	
	First we consider momentum conservation identities. Consider the lattice-sum representation (\ref{eq: mgfdef}) of an MGF with total modular weight $|A|+|B|$ odd (such that it vanishes). Now for each vertex $j\in V_\Gamma$, we can insert $\sum_{e'\in E_\Gamma}\Gamma_{je'} p_{e'}$ (or its complex conjugate) in the lattice sum. This yields \cite{DHoker:2016mwo,Gerken:2020aju}
	\begin{align}
		0&=\sum_{e'\in E_\Gamma}\Gamma_{je'}\sum_{p_e}^\prime \prod_{e\in E_\Gamma}\frac{p_{e'}}{p_e^{a_e}\Bar{p}_e^{b_e}}\prod_{i\in V_\Gamma}\delta\left(\sum_{e''\in E_\Gamma}\Gamma_{ie'' }p_{e''}\right)\,,\nonumber\\0&=\sum_{e'\in E_\Gamma}\Gamma_{je'}\sum_{p_e}^\prime \prod_{e\in E_\Gamma}\frac{\Bar{p}_{e'}}{p_e^{a_e}\Bar{p}_e^{b_e}}\prod_{i\in V_\Gamma}\delta\left(\sum_{e''\in E_\Gamma}\Gamma_{ie'' }p_{e''}\right)\,.
		\label{eq: momconservlattice}
	\end{align}
	Even though we changed the total weight from odd to even by inserting one power of momentum, the expressions still vanish because of the Kronecker deltas. By canceling the factor of momentum in the numerator with one in the denominator, we can write these identities purely in terms of manipulations of edge labels
	\begin{align}
		0=\sum_{e\in E_\Gamma}\Gamma_{je}\mathcal{C}_{\Gamma_{a_e\rightarrow a_e-1}}=\sum_{e\in E_\Gamma}\Gamma_{je}\mathcal{C}_{\Gamma_{b_e\rightarrow b_e-1}}\,,
		\label{eq: momconservshort}
	\end{align}
	for any $j\in V_\Gamma$, which renders the expressions in (\ref{eq: momconservlattice}) as identities between distinct non-vanishing MGFs. In terms of the block notation for dihedral MGFs, this can be written as \cite{DHoker:2016mwo}
	\begin{align}
		0=\sum_{e=1}^r\cno{A-S_e}{B}=\sum_{e=1}^r\cno{A}{B-S_e}\,,
		\label{eq: momconservdihedral}
	\end{align}
	with straightforward generalizations to graphs with more vertices. The reason these are essential to the algorithm is that in some cases they can get rid of negative lower entries, e.g. 
	\begin{equation}
		\cformtri{1\\-1}{2\\1}{2\\1}=\cformtri{1\\0}{2\\0}{2\\1}\,,
		\label{eq: momconservexample}
	\end{equation}
	the significance of which will be explained in section \ref{subsection: trees}. Note that this procedure can produce divergent MGFs, as it changes the sum of the labels of a particular edge and thereby the powers of its inverse momenta in the lattice sums. However, if we start with a convergent MGF and apply a momentum conservation identity, then the possible divergences of individual terms must cancel when you add the results \cite{Gerken:2020aju}. 
	
	The next important tool is holomorphic subgraph reduction. This technique of producing identities between MGFs was first introduced in \cite{DHoker:2016mwo} for dihedral MGFs and was later extended to higher-vertex graphs in \cite{Gerken:2018zcy}. We will review the general idea of HSR and apply it to dihedral graphs as an example. For closed-form expressions of higher-vertex generalizations we refer to \cite{DHoker:2016mwo,Gerken:2018zcy}. The basic idea of HSR relies on partial fraction decomposition of closed holomorphic subgraphs, i.e. subgraphs of MGFs that form a loop of which all edges have anti-holomorphic label zero. For a closed dihedral holomorphic subgraph, we can isolate their momenta and perform partial fraction decomposition as follows
	\begin{align}
		\frac{1}{p^a(q-p)^b}=\sum_{k=1}^a\binom{a+b-k-1}{a-k}\frac{1}{p^kq^{a+b-k}}+\sum_{k=1}^b\binom{a+b-k-1}{b-k}\frac{1}{q^{a+b-k}(q-p)^k}\,.
		\label{eq: partialfrac}
	\end{align}
	We are now in the position to perform the implicit sum over the loop momentum $q$ explicitly\footnote{This sum might only be conditionally convergent for certain values of $a,b$. In these cases we supply the Eisenstein summation prescription. This will break the modular properties of individual terms, but the total result will still be modular.}, meaning the MGF will have one momentum less and therefore one edge less. 
	
	For a general dihedral graph with a closed holomorphic subgraph, we can write the result in closed form 
	\begin{align}
		\cforml{a_+&a_-&A\\0&0&B}&=(-1)^{a_+}\GG_{a_0}\cforml{A\\B}-\binom{a_0}{a_-}\cforml{a_0&A\\0&B}+\sum_{k=4}^{a_+}\binom{a_0-1-k}{a_+-k}\GG_k\cforml{a_0-k&A\\0&B}\nonumber\\&+\sum_{k=4}^{a_-}\binom{a_0-1-k}{a_--k}\GG_k\cforml{a_0-k&A\\0&B}+\binom{a_0-2}{a_+-1}\left\{\hat{\GG}_2\cforml{a_0-2&A\\0&B}+\frac{\pi}{\tau_2}\cforml{a_0-1&A\\-1&B}\right\}\,,
		\label{eq: dihedralHSR}
	\end{align}
	where $a_0=a_++a_-\geq 3$ for absolute convergence. We see the appearance of the non-holomorphic modular form $\hat{\GG}_2$ here, which is defined as $\hat{\GG}_2=\GG_2-\tfrac{\pi}{\tau_2}$\footnote{Here $\GG_2$ does not refer to the $k=2$ instance of (\ref{eq: holEis}), as it is only conditionally convergent. By using the Eisenstein summation prescription we can make it converge and this is what $\GG_2$ refers to. However, this prescription breaks modularity, so we subtract the non-holomorphic term $\tfrac{\pi}{\tau_2}$ to restore it.}. Note that taking the complex conjugate of (\ref{eq: dihedralHSR}) gives a formula to reduce anti-holomorphic subgraphs in terms of $\overline{\GG}_k$ and $\hat{\overline{\GG}}_2$.
	
	It is also possible to write down a closed form for trihedral HSR \cite{DHoker:2016mwo}, but for higher-vertex graphs this becomes rather cumbersome \cite{Gerken:2018zcy}. In fact, HSR is a special case of a more general set of identities called Fay identities \cite{Fay}. These Fay identities hold for any two adjacent holomorphic edges and can be represented graphically as \cite{Gerken:2020aju}    
	
	\begin{align}
		\begin{tikzpicture}[baseline=(zero.base),scale=0.55]
			\node (1) at (0,0)  {$1$};
			\node (2) at (2.25,3.032)  {$2$};
			\node (3) at (4.5,0)  {$3$};
			\node (zero) at (2.25,1.516){};
			\path (2) --node[pos=0.33,vertexdot](4){} node[pos=0.66,vertexdot](5){} (3);
			\draw[->-] (5)--(3);
			\draw[->-] (2)--(4);
			\draw[loosely dotted,thick,dash phase=1pt] (4)--(5);
			\draw[->-] (1)--node[label,pos=0.4,fill=white]{$\scriptstyle(a_{1}{,}0)$}(2);
			\draw[->-] (1)--node[label,pos=0.5,fill=white]{$\scriptstyle(a_{2}{,}0)$}(3);
		\end{tikzpicture}
		&=(-1)^{a_{1}}
		\begin{tikzpicture}[baseline=(2.base),scale=0.55]
			\node (2) at (0,0)  {$2$};
			\node (3) at (4.5,0)  {$3$};
			\path (2) to[bend left=60] node[pos=0.33,vertexdot](4){} node[pos=0.66,vertexdot](5){} (3);
			\draw[->-] (2) to [bend left=23](4);
			\draw[->-] (5) to [bend left=23](3);
			\draw[loosely dotted,thick,dash phase=1pt] (4) to [bend left=5](5);
			\draw[->-]
			(2)--node[label,pos=0.44]{$\scriptstyle(a_{1}{+}a_{2}{,}0)$}(3);
		\end{tikzpicture}\nonumber\\[-1em]
		&\hspace{-1em}-\binom{a_{1}{+}a_{2}{-}1}{a_{1}}
		\begin{tikzpicture}[baseline=(zero.base),scale=0.55]
			\node (1) at (0,0)  {$1$};
			\node (2) at (2.25,3.032)  {$2$};
			\node (3) at (4.5,0)  {$3$};
			\node (zero) at (30:2.021){};
			\path (2) --node[pos=0.33,vertexdot](4){} node[pos=0.66,vertexdot](5){} (3);
			\draw[->-] (2)--(4);
			\draw[->-] (5)--(3);
			\draw[loosely dotted,thick,dash phase=1pt] (4)--(5);
			\draw[->-]
			(1)--node[label,pos=0.44]{$\scriptstyle(a_{1}{+}a_{2}{,}0)$}(3);
		\end{tikzpicture}
		-\binom{a_{1}{+}a_{2}{-}1}{a_{2}}
		\begin{tikzpicture}[baseline=(zero.base),scale=0.55]
			\node (1) at (0,0)  {$1$};
			\node (2) at (2.25,3.032)  {$2$};
			\node (3) at (4.5,0)  {$3$};
			\node (zero) at (30:2.021){};
			\path (2) --node[pos=0.33,vertexdot](4){} node[pos=0.66,vertexdot](5){} (3);
			\draw[->-] (2)--(4);
			\draw[->-] (5)--(3);
			\draw[loosely dotted,thick,dash phase=1pt] (4)--(5);
			\draw[->-]
			(1)--node[label,pos=0.4]{$\scriptstyle(a_{1}{+}a_{2}{,}0)$}(2);
		\end{tikzpicture}\label{eq: Fay}\nonumber\\
		&\hspace{-1em}+\sum_{j=0}^{a_{1}-1}\binom{a_{2}{+}j{-}1}{j}
		\begin{tikzpicture}[mgf,baseline=(zero.base),scale=0.55]
			\node (1) at (0,0)  {$1$};
			\node (2) at (2.25,3.032)  {$2$};
			\node (3) at (4.5,0)  {$3$};
			\node (zero) at (30:2.021){};
			\path (2) to[bend left=60] node[pos=0.33,vertexdot](4){} node[pos=0.66,vertexdot](5){} (3);
			\draw[->-] (2) to [bend left=23](4);
			\draw[->-] (5) to [bend left=23](3);
			\draw[loosely dotted,thick,dash phase=1pt] (4) to [bend left=5](5);
			\draw[->-]
			(3)--node[label,pos=0.4,inner sep=1pt]{$\scriptstyle(a_{1}{-}j{,}0)$}(2);
			\draw[->-]
			(1)--node[label,pos=0.45]{$\scriptstyle(a_{2}{+}j{,}0)$}(3);
		\end{tikzpicture}
		+\sum_{j=0}^{a_{2}-1}\binom{a_{1}{+}j{-}1}{j}
		\begin{tikzpicture}[baseline=(zero.base),scale=0.55]
			\node (1) at (0,0)  {$1$};
			\node (2) at (2.25,3.032)  {$2$};
			\node (3) at (4.5,0)  {$3$};
			\node (zero) at (30:2.021){};
			\path (2) to[bend left=60] node[pos=0.33,vertexdot](4){} node[pos=0.66,vertexdot](5){} (3);
			\draw[->-] (2) to [bend left=23](4);
			\draw[->-] (5) to [bend left=23](3);
			\draw[loosely dotted,thick,dash phase=1pt] (4) to [bend left=5](5);
			\draw[->-]
			(1)--node[label,pos=0.45]{$\scriptstyle(a_{1}{+}j{,}0)$}(2);
			\draw[->-]
			(2)--node[label,pos=0.55,inner sep=1pt]{$\scriptstyle(a_{2}{-}j{,}0)$}(3);
		\end{tikzpicture}\,,
	\end{align}
	where the ellipses denote a sequence of edges. Whenever this sequence of edges is purely holomorphic, we indeed see it allows us to convert $n$-point closed holomorphic subgraphs to $(n-1)$-point closed holomorphic subgraphs plus graphs of lower loop order (whenever we take into account the other edges of the MGF). Now by repeatedly applying Fay identities, we can convert a graph containing an $n$-point closed holomorphic subgraph to a linear combination of graphs, some of which may contain a dihedral holomorphic subgraph. As we already wrote the general reduction formula for dihedral HSR in (\ref{eq: dihedralHSR}), we can reduce the original MGF containing the $n$-point closed holomorphic subgraph to a linear combination of MGFs free of holomorphic subgraphs. This is a key point upon which the algorithm is built that will be explained in more detail in the next section.
	
	The modular equivariant iterated integrals that will be introduced in section \ref{section: Iterint} are non-holomorphic modular forms of purely anti-holomorphic weight $(0,b)$. As we want to translate any MGF of modular weights $(|A|,|B|)$ to these equivariant iterated integrals, a different convention for MGFs in (\ref{eq: mgfdef}) such that they are also of purely anti-holomorphic weight turns out to be useful \cite{DHoker:2016mwo}. We do it as follows. For every holomorphic edge label $a_e$ we add a factor of $\tau_2^{a_e}$. As $\tau_2$ is a non-holomorphic modular form of weight $(-1,-1)$ this changes an MGF of weights $(|A|,|B|)$ to one of weights $(0,|B|-|A|)$, so that we cancel all holomorphic weights. Furthermore, for all edge labels $a_e$ and $b_e$ we add factors of $\pi^{-a_e/2}$ and $\pi^{-b_e/2}$ respectively, so that for $\mathcal{C}_\Gamma(\tau)$ defined in (\ref{eq: mgfdef}) of weights $(|A|,|B|)$, we have 
	\begin{align}
		\mathcal{C}^+_\Gamma(\tau):=\frac{\tau_2^{|A|}}{\pi^{(|A|+|B|)/2}}\mathcal{C}_\Gamma(\tau)\,.
		\label{eq: cplus}
	\end{align}
	This notation naturally carries over to equations (\ref{eq: dihedralC}) and (\ref{eq: trihedralC}).
	
	In (\ref{eq: maassact}) we saw that the holomorphic differential operator transformed modular forms of weights $(a,b)$ into modular forms of weights $(a+1,b-1)$. In particular, it would send the newly defined MGFs in (\ref{eq: cplus}) to MGFs of weights $(1,|B|-|A|-1)$. However, we would like to stay within the space of MGFs of purely anti-holomorphic weight in view of the iterated integrals. Therefore we define a variant of (\ref{eq: Maass})\footnote{Note that we only redefine the holomorphic differential operator here. We can redefine the anti-holomorphic one is similar fashion, but it will play no role in the algorithm. It is possible to use the anti-holomorphic operator to define the algorithm instead, but this will not generate additional results as we can switch holomorphic with anti-holomorphic labels by complex conjugation.} as \cite{DHoker:2016mwo}
	\begin{align}
		\pi\nabla_0:=\pi \tau_2\nabla^{(0)}:(0,b)\rightarrow (0,b-2)\,.
		\label{eq: pinabla}
	\end{align}
	In particular, we have 
	\begin{align}
		\pi\nabla_0\mathcal{C}^+_\Gamma(\tau)=\pi\sum_{e\in E_\Gamma}a_e\mathcal{C}^+_{\Gamma_{(a_e,b_e)\rightarrow (a_e+1,b_e-1)}}(\tau)\,.
		\label{eq: pinablaact}
	\end{align}
	All other identities such as momentum conservation and HSR might take slightly different forms in this convention due to the extra powers of $\tau_2$ and $\pi$, but the ideas remain exactly the same.

	\section{Constructing tree-representations of MGFs}
	\label{subsection: trees}
	The backbone of the conversion algorithm that translates lattice-sum representations of MGFs into Equivariant Iterated Eisenstein Integrals (EIEIs) are the tree-representations of MGFs we can construct using the differential operators and identities introduced in sections \ref{subsection: mgfintro} and \ref{subsec: identities}. However, the definition of an MGF for any directed labeled simple graph $\Gamma$ given by (\ref{eq: mgfdef}) is too broad for our purposes. In fact, the space spanned by these MGFs is not contained in the one spanned by the EIEIs. This means that not every MGF defined by (\ref{eq: mgfdef}) can be written in terms of EIEIs. As the purpose of the algorithm is exactly this translation procedure, it makes sense to restrict our input so that this procedure is well-defined.
	
	We can identify the anomalous MGFs that are not amenable to the translation procedure. In fact, there are two specific classes of MGFs that are outside the space of EIEIs. First, the iterated integrals to be introduced in section \ref{section: Iterint} are regularized, while MGFs can be divergent, hence we reasonably restrict the input to convergent MGFs. Second, if we start with an MGF containing a (anti-)holomorphic subgraph, iterative HSR will produce a linear combination of MGFs free of (anti-)holomorphic subgraphs, some of which are multiplied by factors of $\GG_{k_i}$ or $\hat{\GG}_2$ as a result of (\ref{eq: dihedralHSR}) (or their complex conjugates). As $\GG_{k_i}$ and $\hat{\GG}_2$ (and their complex conjugates) cannot be written as EIEIs by definition, the same holds for the original MGF. Nonetheless, all the remaining MGFs including the coefficients of the Eisenstein series are within the space spanned by EIEIs. Hence, the original MGF can be written as a linear combination of EIEIs and (anti-)holomorphic Eisenstein series with EIEI coefficients, but the full expression and thus the original MGF is outside of the space of EIEIs.
	
	The exclusion of these two types of MGFs: divergent MGFs and the ones containing (anti-)holomorphic subgraphs, is enough to specify the space of convertible MGFs. Another way to restrict to this subspace immediately is to only consider MGFs that arise from the one-loop closed-string amplitude generating series \cite{Gerken:2019cxz,Gerken:2020yii}, which is conjecturally the generating series of all one-loop closed strings integrands (modulo theory-dependent partition functions).
	
	Still, as we will see, for us to be able to write down a tree for an MGF, we need  $a_e,b_e\geq 0$ for all edge labels. It is easy to see that if the MGF is convergent and free of holomorphic and anti-holomorphic subgraphs, it can always be written as a linear combination of MGFs of this form using momentum conservation identities. We can then perform the conversion for every single MGF in the linear combination and add the results.
	
	Now, let us describe how to construct a tree for a given MGF of the form described above. First of all, we change conventions to $(\ref{eq: cplus}$) so that we have an MGF of modular weights $(0,|B|-|A|)$. Applying the holomorphic differential operator (\ref{eq: pinabla}) to this MGF turns it into a linear combination of MGFs with modular weights $(0,|B|-|A|-2)$. Now one of the following things can happen for every individual MGF in the linear combination:
	\begin{enumerate}
		\item $b_e\geq 0$ for all $e$ and there are no holomorphic subgraphs except for possible factors of $\GG_k$.
		\item At least one of the $b_e=-1$.
		\item We have an $n$-point closed holomorphic subgraph.
	\end{enumerate}
	We now look at every MGF in the linear combination case by case and see where it lands in this scheme. As the statements are mutually exclusive, this splits the result in three separate sets, some of which can be empty.
	
	The next step is to apply the techniques of momentum conservation and HSR to the MGFs depending on which set they belong to. The logic only works if we treat the sets in an orderly fashion. For the MGFs that landed in the first set, we do nothing. To all MGFs that landed in the second set, we apply momentum conservation identities to remove the negative $b_e$. This in turn creates a linear combination of different MGFs, which we have to assign to one of the sets again. As statement 2 will now be false, they get assigned to either set 1 or 3. If they land in set 1, we leave them be. Otherwise we adjoin them to set 3. For all the MGFs in set 3, we iteratively apply Fay identities to the $n$-point closed holomorphic subgraphs until we reach dihedral closed holomorphic subgraphs, after which we apply (\ref{eq: dihedralHSR}). 
	
	It turns out that in this series of steps, MGFs that require HSR always come in pairs such that we can cancel the last two terms of (\ref{eq: dihedralHSR}) \cite{DHoker:2016mwo}. For example
	\begin{align}
		\cforml{2\,2\,A\\0\,0\,B}-2\cforml{3\,1\,A\\0\,0\,B}=2\cforml{4\,A\\0\,B}+3\GG_4\cforml{A\\B}\,.
		\label{eq: dihedralHSRexample}
	\end{align}
	This shows that dihedral HSR reduces MGFs containing dihedral closed holomorphic subgraphs to a linear combination of MGFs without closed holomorphic subgraphs, some of which are multiplied by factors of $\GG_k$. For these MGFs statement 1 is true while 2 and 3 are false, which is now the case for all MGFs.
	
	After performing this sequence of steps, we end up with a linear combination of convergent MGFs free of closed holomorphic subgraphs (except for possible factors of $\GG_k$) and free of negative $b_e$. This means we are almost in the same position as we started with, but now with several MGFs of weights $(0,|B|-|A|-2)$. The fact that we might have factors of holomorphic Eisenstein series now prevents us from applying the above steps again to all the MGFs of weight $(0,|B|-|A|-2)$, as applying the differential operator to $\GG_k$ creates a negative $b_e$ that cannot be removed using momentum conservation identities. Therefore, whenever we encounter a $\GG_k$ multiplying another MGF, we remove it manually along with $k$ factors of $\tau_2$ and separate these MGFs from the others and group them into sets depending on the value of $k$ in $\GG_k$. Now we are in the position to apply all the steps again.
	
	We can visualize a step in this process graphically, which turns out to be very useful. Let $\Gamma$ denote the convergent MGF we start with of weights $(0,|B|-|A|)$. We act with the differential operator and go through the steps outlined above. In the end we will have several sets of MGFs left over; a set of MGFs that did not multiply any $\GG_k$ and sets that did, separated by the value of $k$. We visualize this splitting by drawing corresponding edges going downward to vertices where the vertices are accompanied by the $\GG_k$ (and $k$ factors of $\tau_2$ that we do not explicitly write)
	\begin{align}
		\begin{tikzpicture}[baseline=(zero.base)]
			\draw (0,1) to (-2,0);
			\draw (0,1) to (-1,0);
			\draw (0,1) to (2,0);
			\node[fill=white] at (0,1) {$\Gamma$};    
			\fill (-2,0) circle (2pt);
			\fill (-1,0) circle (2pt);
			\fill (2,0) circle (2pt);
			\node[right] at (-0.95,0) {$\GG_{k_1}$};
			\node[right] at (2,0) {$\GG_{k_n}$};
			\node[right] at (0,1/4) (zero) {$\boldsymbol{\cdots}$};
		\end{tikzpicture}\,,
		\label{eq: singlestep}
	\end{align}
	where the leftmost branch represents the MGFs of weights $(0,|B|-|A|-2)$ not multiplying any $\GG_{k_i}$. For the other branches, removing factors of $\tau_2^{k_i} \GG_{k_i}$ removes modular weights $(0,-k_i)$, such that the MGFs at those branches are of modular weights $(0,|B|-|A|-2+k_i)$.
	
	The black vertices represent the MGFs that multiply the corresponding $\GG_{k_i}$. These MGFs will be important for the next step of the process, but they will not matter once we build the entire tree, hence the reason why we represent them by the same black vertex. However, this is not true whenever the black vertex represents a modular graph function, i.e. an MGF of weights $(0,0)$\footnote{The conditions for this to happen are the following. As $|A|+|B|$ is even, so is the anti-holomorphic weight $|B|-|A|$. We saw that for each step we go down the tree we change the anti-holomorphic weight by $-2$ or by $-2+k_i$, which are also even numbers. Then, whenever $|B|-|A|-2n+\sum_{i\in I} k_i=0$ is satisfied at a certain vertex, with $n$ the amount of steps down the tree to that vertex and $I$ the index set of the $\GG_{k_i}$ encountered along the way, we will have modular graph functions.}. Whenever this happens, we will see that we need to remember the particular modular graph functions as these will fix the uniqueness of the iterated integral representation of the original MGF. This will be explained in more detail in section \ref{subsection: constants}. To distinguish the vertices that contain modular graph functions from the ones that contain modular graph forms, we color them orange (similarly if the $\Gamma$ we started with was modular invariant, we color $\Gamma$ orange).
	
	Now, applying the steps again to all the MGFs at the lower vertices of (\ref{eq: singlestep}), we get a tree that terminates after $N=|B|$ iterations. This is because we keep producing closed holomorphic subgraphs that will yield factors of $\GG_k$ upon HSR. Therefore, the sum of all $k_i$ of $\GG_{k_i}$ along any branch must be equal to $|A|+|B|$. After $|B|$ iterations, all the leftover $b_e$ are equal to zero, so the only thing left will be some rational numbers multiplying some holomorphic Eisenstein series. These rational numbers are very important in the conversion to iterated integrals, hence we represent the lowest vertices by an open vertex to distinguish them from the black and orange vertices. Now let us look at some concrete examples.
	
	\subsection{Examples of trees}
	
	First we start with one of the easiest nontrivial cases, which is the dihedral graph $\Gamma=\cformp{2\,2\,1\\2\,1\,1}$. The tree will terminate after $|B|=4$ iterations of the outlined process. Besides the differential operation, we will draw the corresponding tree where we add a new layer in red for every iteration. Further, note that $\Gamma$ is modular invariant, so according to the rules we color it orange. Applying the differential operator ($\ref{eq: pinabla}$) once gives
	\begin{align}
		\pi\nabla_0 \cformp{2\,1\,1\\2\,1\,1}=2\pi\cformp{3\,1\,1\\1\,1\,1}+2\pi\cformp{2\,2\,1\\2\,0\,1}\quad\longleftrightarrow\quad 
		\begin{tikzpicture}[baseline=(zero.base)]
			\draw[color=red] (0,0) to (0,-1);
			\fill[color=red] (0,-1) circle (2pt);
			\node at (0,-3/4) (zero) {};
			\node[fill=white] at (0,0) {\textcolor{orange}{$\cformp{2\,1\,1\\2\,1\,1}$}};
		\end{tikzpicture}\,.
		\label{eq: firstiter}
	\end{align}
	As all the MGFs on the Right-Hand Side (RHS) of the equality are free of negative $b_e$, free of closed holomorphic subgraphs and there is no $\GG_k$, we apply another $\pi\nabla_0$. This yields
	\begin{align}
		\pi^2\nabla_0^2\cformp{2\,1\,1\\2\,1\,1}=6\pi^2\cformp{4\,1\,1\\0\,1\,1}+8\pi^2\cformp{3\,2\,1\\1\,0\,1}+4\pi^2\cformp{2&3&1\\2&-1&1}+2\pi^2\cformp{2\,2\,2\\2\,0\,0}\,.
		\label{eq: seconditerprimitive}
	\end{align}
	The first two MGFs on the RHS land in set 1. The third one lands in set 2, while the last one lands in set 3. Following the steps, we apply momentum conservation to the third one, which gives
	\begin{align}
		\cformp{2&3&1\\2&-1&1}=-\cformp{2\,3\,1\\2\,0\,0}-\cformp{2\,3\,1\\1\,0\,1}\,,
		\label{eq: momconservex}
	\end{align}
	of which the first on the RHS lands in set 3, while the second one lands in set 1. Now we can substitute this into equation (\ref{eq: seconditerprimitive}) and apply dihedral HSR (\ref{eq: dihedralHSR}) to all the dihedral closed holomorphic subgraphs. Finally, this gives
	\begin{align}
		\pi^2\nabla_0^2\cformp{2\,1\,1\\2\,1\,1}=6\pi^2\cformp{4\,1\,1\\0\,1\,1}+8\pi^2\cformp{3\,2\,1\\1\,0\,1}-4\pi^2\cformp{2\,3\,1\\1\,0\,1}+4\pi^2\cformp{6\,0\\2\,0}\,\longleftrightarrow\, 
		\begin{tikzpicture}[baseline=(zero.base)]
			\draw (0,0) to (0,-1);
			\draw[color=red] (0,-1) to (0,-2);
			\fill (0,-1) circle (2pt);
			\fill[color=red] (0,-2) circle (2pt);
			\node at (0,-5/4) (zero) {};
			\node[fill=white] at (0,0) {\textcolor{orange}{$\cformp{2\,1\,1\\2\,1\,1}$}};
		\end{tikzpicture}\,.
		\label{eq: seconditer}
	\end{align}
	All MGFs on the RHS are assigned to set 1 now and we see no $\GG_k$, so we continue once more. This generates a large amount of MGFs, but after applying momentum conservation and performing dihedral HSR, we are left with
	\begin{align}
		\pi^3\nabla_0^3 \cformp{2\,1\,1\\2\,1\,1}=108\pi^3\cformp{7\,0\\1\,0}-12\pi\cformp{3\,0\\1\,0}\tau_2^4\GG_4\quad\longleftrightarrow\quad 
		\begin{tikzpicture}[baseline=(zero.base)]
			\draw (0,0) to (0,-1);
			\draw (0,-1) to (0,-2);
			\draw[color=red] (0,-2) to (-1,-3);
			\draw[color=red] (0,-2) to (1,-3);
			\fill (0,-1) circle (2pt);
			\fill (0,-2) circle (2pt);
			\fill[color=red] (-1,-3) circle (2pt);
			\fill[color=red] (1,-3) circle (2pt);
			\node at (0,-7/4) (zero) {};
			\node[fill=white] at (0,0) {\textcolor{orange}{$\cformp{2\,1\,1\\2\,1\,1}$}};
			\node[right,color=red] at (1,-3) {$\GG_4$};
		\end{tikzpicture}\,.
		\label{eq: thirditer}
	\end{align}
	The first MGF on the RHS has no factor of $\GG_k$, while the second one is multiplied with $\GG_4$. We separate this $\GG_4$ along with $\tau_2^4$ from the accompanying MGF and leave it at a separate branch of the tree. We now apply the last iteration of the process, after which it should terminate. These give
	\begin{align}
		\pi\nabla_0 \left(108\pi^3\cformp{7\,0\\1\,0}\right)=756\tau_2^8\GG_8,\quad \pi\nabla_0\left(-12\pi\cformp{3\,0\\1\,0}\right)=-36\tau_2^4\GG_4\,.
		\label{eq: fourthiter}    
	\end{align}
	Indeed we see the process terminates and the only things that remain are rational numbers multiplying holomorphic Eisenstein series. The full tree therefore is
	
	
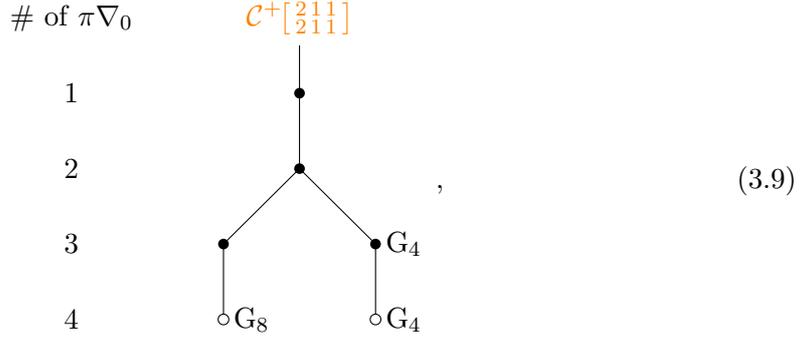
\begin{align}
		\begin{tikzpicture}[baseline=(zero.base)]
			\draw (0,0) to (0,-1);
			\draw (0,-1) to (0,-2);
			\draw (0,-2) to (-1,-3);
			\draw (0,-2) to (1,-3);
			\draw (-1,-3) to (-1,-4);
			\draw (1,-3) to (1,-4);
			\fill (0,-1) circle (2pt);
			\fill (0,-2) circle (2pt);
			\fill (-1,-3) circle (2pt);
			\fill (1,-3) circle (2pt);
			\node at (0,-9/4) (zero) {};
			\node[fill=white] at (0,0) {\textcolor{orange}{$\cformp{2\,1\,1\\2\,1\,1}$}};
			\node[right] at (1,-3) {$\GG_4$};
			\draw[fill=white] (-1,-4) circle[radius=2pt];
			\draw[fill=white] (1,-4) circle[radius=2pt];
			\node[right] at (1,-4) {$\GG_4$};
			\node[right] at (-1,-4) {$\GG_8$};
			\node[fill=white] at (-3,0) {$\#$ of $\pi\nabla_0$};
			\node[fill=white] at (-3,-1) {$1$};
			\node[fill=white] at (-3,-2) {$2$};
			\node[fill=white] at (-3,-3) {$3$};
			\node[fill=white] at (-3,-4) {$4$};
		\end{tikzpicture}\,,
		\label{eq: fulltree}
	\end{align}
	where the left and right open vertices represent the numbers 672 and -36 respectively. Note that nowhere along the process did we encounter modular graph functions except the fact that the MGF we started with was one.
	
	It turns out that this graphical representation, along with the precise values of the numbers at the bottom and the knowledge of the constants in the cuspidal expansions of modular graph functions that appeared in the process, completely and uniquely represent the MGFs in terms of iterated integrals. This will be shown in section \ref{section: alg}.
	
	To end this section, we show some more examples of trees that terminate after 4 iterations
	
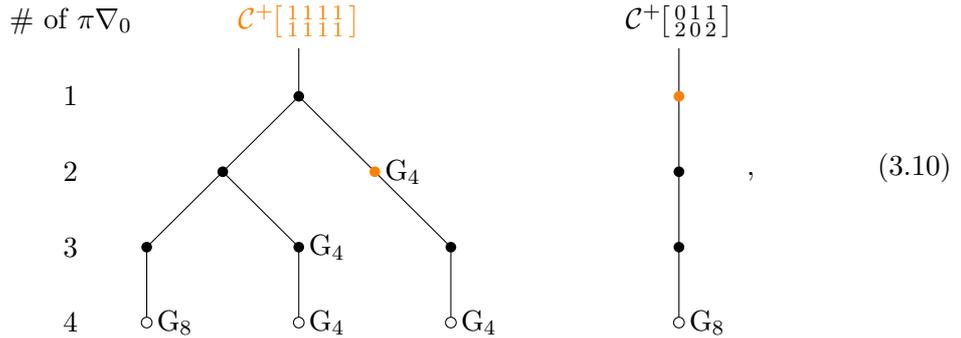
\begin{align}
		\begin{tikzpicture}[baseline={([yshift=-0.5ex]current bounding box.center)}]
			\draw[] (1,1) to (1,0); 
			\draw[] (1,0) to (0,-1);
			\draw[] (1,0) to (2,-1);
			\draw[] (0,-1) to (-1,-2);
			\draw[] (0,-1) to (1,-2);
			\draw[] (2,-1) to (3,-2);
			\draw[] (-1,-2) to (-1,-3);
			\draw[] (1,-2) to (1,-3);
			\draw[] (3,-2) to (3,-3);
			\draw[] (6,1) to (6,0);
			\draw[] (6,0) to (6,-1);
			\draw[] (6,-1) to (6,-2);
			\draw[] (6,-2) to (6,-3);
			\fill (-1,-2) circle (2pt);
			\fill (1,-2) circle (2pt);
			\fill (3,-2) circle (2pt);
			\fill (1,0) circle (2pt);
			\fill (0,-1) circle (2pt);
			\fill[color=orange] (2,-1) circle (2pt);
			\fill[color=orange] (6,0) circle (2pt);
			\fill (6,-1) circle (2pt);
			\fill (6,-2) circle (2pt);
			\draw[fill=white] (-1,-3) circle[radius=2pt];
			\draw[fill=white] (1,-3) circle[radius=2pt];
			\draw[fill=white] (3,-3) circle[radius=2pt];
			\draw[fill=white] (6,-3) circle[radius=2pt];
			\node[right] at (-1,-3) {$\GG_8$};   
			\node[right] at (1,-3) {$\GG_4$}; 
			\node[right] at (3,-3) {$\GG_4$};
			\node[right] at (2,-1) {$\GG_4$};
			\node[right] at (1,-2) {$\GG_4$};
			\node[right] at (6,-3) {$\GG_8$};
			\node[fill=white] at (1,1) {\textcolor{orange}{$\cformp{1\,1\,1\,1\\1\,1\,1\,1}$}};    
			\node[fill=white] at (-2,1) {$\#$ of $\pi\nabla_0$};
			\node[fill=white] at (-2,0) {$1$};
			\node[fill=white] at (-2,-1) {$2$};
			\node[fill=white] at (-2,-2) {$3$};
			\node[fill=white] at (-2,-3) {$4$};
			\node[fill=white] at (6,1) {$\cformp{0\,1\,1\\2\,0\,2}$};   
		\end{tikzpicture}\,,
		\label{eq: othertrees}
	\end{align}
	where $\cformp{1\,1\,1\,1\\1\,1\,1\,1}$ is commonly known as $D_4$ or the three-loop banana graph in the literature.
	
	As we have seen, these tree-representations of MGFs are built using the iterative action of certain differential operators. Then, to represent the MGF in terms of iterated integrals, we would like to integrate the result back up. The particular way to do this is encoded in the structure of the tree. Furthermore, we see that the objects we need to integrate back up, i.e. the integration kernels, will involve the holomorphic Eisenstein series $\GG_k$. Also, as the MGFs we would like to represent using iterated integrals are modular, we are looking for modular iterated integrals of holomorphic Eisenstein series. We discuss this particular type of iterated integrals in the next section and integrate back up in section \ref{section: alg}.

	\section{Review of equivariant iterated Eisenstein integrals}
	Infinite families of non-holomorphic modular forms that arise from iterated integrals of holomorphic Eisenstein series and their complex conjugates were constructed in \cite{Brown:2017qwo,Brown:2017qwo2,Broedel:2018izr,Gerken:2019cxz,Gerken:2020yii,drewitt2022laplace,Dorigoni:2022npe,Dorigoni:2024oft}. These are the coefficients of certain generating series called equivariant iterated Eisenstein integrals (EIEIs), which contain MGFs \cite{Brown:2017qwo,Brown:2017qwo2,Dorigoni:2022npe, Dorigoni:2024oft}. We will not introduce the entire construction of EIEIs, but review the relevant parts for this work and refer to \cite{Brown:2017qwo,Brown:2017qwo2,Broedel:2018izr,Gerken:2019cxz,Gerken:2020yii,drewitt2022laplace,Dorigoni:2022npe,Dorigoni:2024oft} for details. The dictionary between MGFs and the equivariant iterated integrals has been under construction since the advent of EIEIs \cite{Brown:2017qwo,drewitt2022laplace,Gerken:2019cxz,Gerken:2020yii,Dorigoni:2021jfr,Dorigoni:2021ngn,Dorigoni:2024oft,Dorigoni:2022npe,Broedel:2018izr} and a systematic algorithm to translate any convergent MGF to its representation in terms of iterated integrals will be outlined in section \ref{section: alg}. In this section we review the relevant iterated integrals and the differential equation they satisfy. We follow \cite{Dorigoni:2022npe,Dorigoni:2024oft}.
	\label{section: Iterint}
	
	\subsection{Equivariant iterated integrals of holomorphic modular forms}
	\label{subsection: kernels}
	To construct iterated integrals of holomorphic Eisenstein series, we need a set of integration kernels. A particularly convenient choice of kernels is the following
	\begin{align}
		\omplus{j}{k}{\tau,\tau_1}&=\frac{d\tau_1}{2\pi i}\left(\frac{\tau-\tau_1}{4y}\right)^{k-2-j}\left(\Bar{\tau}-\tau_1\right)^j\GG_k(\tau_1)\,,\nonumber\\\omminus{j}{k}{\tau,\tau_1}&=-\frac{d\Bar{\tau}_1}{2\pi i}\left(\frac{\tau-\Bar{\tau}_1}{4y}\right)^{k-2-j}\left(\Bar{\tau}-\Bar{\tau}_1\right)^j\overline{\GG_k(\tau_1)}\,,
		\label{eq: kernels}
	\end{align}
	with $k\geq 4$ even and $0\leq j\leq k-2$. These are engineered such that they transform as modular forms of modular weights $(0, k-2-2j)$ in the following way
	\begin{align}
		\ompm{j}{k}{\frac{\alpha\tau+\beta}{\gamma\tau+\delta},\frac{\alpha\tau_1+\beta}{\gamma\tau_1+\delta}} = (\gamma\bar\tau+\delta)^{k-2-2j}\ompm{j}{k}{\tau,\tau_1}\,.
	\end{align}
	Their iterated integral versions are defined as
	\begin{align}
		\betaplustau{j_1&j_2&\cdots&j_\ell\\k_1&k_2&\cdots&k_\ell}&=\int_\tau^{i\infty}\omplus{j_\ell}{k_\ell}{\tau,\tau_\ell}\dots\int_{\tau_3}^{i\infty}\omplus{j_2}{k_2}{\tau,\tau_2}\int_{\tau_2}^{i\infty}\omplus{j_1}{k_1}{\tau,\tau_1}\,,\nonumber\\\betaminustau{j_1&j_2&\cdots&j_\ell\\k_1&k_2&\cdots&k_\ell}&=\int_{\Bar{\tau}}^{-i\infty}\omminus{j_\ell}{k_\ell}{\tau,\tau_\ell}\dots\int_{\Bar{\tau}_3}^{-i\infty}\omminus{j_2}{k_2}{\tau,\tau_2}\int_{\Bar{\tau}_2}^{-i\infty}\omminus{j_1}{k_1}{\tau,\tau_1}\,,
		\label{eq: betapm}
	\end{align}
	which are homotopy invariant iterated integrals \cite{Dorigoni:2022npe} and regularized by the tangential basepoint regularization \cite{Brown:2014pnb}. The sum $k_1+\dots+k_\ell$ is called the degree of the iterated integral and $\ell$ is called the depth.
	
	The vector space of weight-$k$ holomorphic modular forms does not only involve $\GG_k$. In fact, the vector space of holomorphic modular forms can be decomposed into a one-dimensional space spanned by $\GG_k$ and a space spanned by holomorphic cusp forms $\mathcal{S}_k$. Denoting with $\Delta_k(\tau)\in\mathcal{S}_k$ an unspecified cusp form, we can construct iterated integrals of cusp forms from the same integration kernels as in (\ref{eq: kernels}), but with $\GG_k$ replaced by a cusp form $\Delta_k$. We denote these with $\ompm{j}{\Delta_k}{\tau,\tau_1}$ and similarly we replace $k_i\mapsto \Delta_{k_i}$ in the definitions of the iterated integrals (\ref{eq: betapm}) to define iterated integrals of cusp forms. As all of these objects are iterated integrals, they satisfy shuffle relations
	\begin{align}
		\beta_{\pm}[X;\tau]\beta_{\pm}[Y;\tau]=\sum_{P\in X\shuffle Y}\beta_{\pm}[P;\tau]\,,
		\label{eq: shuffle}
	\end{align}
	where $X\shuffle Y$ denotes the shuffle product of the words $X$ and $Y$.
	
	Despite the fact that the kernels (\ref{eq: kernels}) transform as modular forms, their iterated integrals in (\ref{eq: betapm}) do not. Nonetheless, it is possible to find modular objects corresponding to the iterated integrals (\ref{eq: betapm}) for given depth $\ell$, which we call their modular completions. In the case of Eisenstein kernels\footnote{For cusp form kernels, the modular completions turn out to be much more involved and only depth one completions have been found \cite{brown2017a}.}, these have been explicitly realized for depth $\leq 3$ and degree $\leq 20$ in \cite{Dorigoni:2022npe,Dorigoni:2024oft} and have been found to contain, among others, iterated integrals of depth $\leq 3$ of both $\GG_k$ and $\Delta_k$, Multiple Zeta Values (MZVs), L-values and new periods. Given such an iterated integral as in (\ref{eq: betapm}) of depth $\ell$, we denote its modular completion with $\beta^{\mathrm{eqv}}$. It has the modular transformation property
	\begin{align}
		\betaeqvx{j_1&j_2&\dots&j_\ell\\k_1&k_2&\dots&k_\ell}{\tfrac{a\tau +b}{c\tau +d}}=\left(\prod_{i=1}^\ell (c\Bar{\tau}+d)^{k_i-2j_i-2}\right)\betaeqvtau{j_1&j_2&\dots&j_\ell\\k_1&k_2&\dots&k_\ell}\,,
		\label{eq: betacovariant}
	\end{align}
	with modular weights $(0,\sum_{i=1}^\ell k_i-2j_i-2)$. Being comprised of iterated integrals, they inherit the shuffle property from (\ref{eq: shuffle}). The construction of the $\beta^{\mathrm{eqv}}$ is unique, except whenever it is modular invariant where it is unique up to a constant \cite{Brown:2017qwo,Dorigoni:2022npe,Dorigoni:2024oft}. In section \ref{subsection: diffeqn}, we will comment on how to fix these constants. 
	
	In section \ref{subsection: trees}, we saw that according to the tree-representations of MGFs, the integration kernels of MGFs only involve holomorphic Eisenstein series $\GG_k$ and no cusp forms $\Delta_k$. This means that the iterated integral descriptions of MGFs will not involve any cusp forms. As general modular completions $\beta^\mathrm{eqv}$ of the iterated Eisenstein integrals can involve cusp forms, we see that the space spanned by $\beta^{\mathrm{eqv}}$ is bigger than the space spanned by the subset of MGFs\footnote{The construction of these equivariant iterated integrals is most conveniently represented in terms of a generating series. The noncommutative bookkeeping variables $\epsilon_k$ satisfy certain relations called Tsunogai-relations \cite{Pollack,Tsunogai,Dorigoni:2022npe,Brown:2017qwo2,Gerken:2020yii,Dorigoni:2021ngn}. These relations cause certain linear combinations of $\beta^{\mathrm{eqv}}$ to drop out from the generating series, which are precisely the combinations that cannot be realized through MGFs. A generalization of this generating series to also accommodate the $\beta^{\mathrm{eqv}}$ that go beyond MGFs was found in \cite{Dorigoni:2024oft}.} defined in section \ref{section: MGFs}. This was investigated in \cite{Dorigoni:2021jfr,Dorigoni:2021ngn,Dorigoni:2024oft,Brown:2017qwo2,Gerken:2020yii}, and as we are only interested in MGFs, we refer to these references for details on modular completions involving cusp forms. In particular, the depth $\ell$ modular completions of the iterated integrals realized through MGFs, are completely expressible in terms of linear combinations of depth $\leq \ell$ iterated Eisenstein integrals, along with rational functions of $\tau,\Bar{\tau}$ with $\QQ[\mathrm{MZV}]$\footnote{Conjecturally single-valued MZVs \cite{Zerbini:2015rss,DHoker:2015wxz,Dorigoni:2024oft}.}-coefficients.
	
	An example of a modular equivariant iterated integral of depth $\ell=2$ (free of cusp forms) is
	\begin{align}
		\betaeqv{2&4\\4&6}&=\betaplus{2&4\\4&6}+\betaminus{4&2\\6&4}+\betaplus{4\\6}\betaminus{2\\4}-\frac{2}{3}\betaplus{4\\6}\zeta_3-\frac{2}{5}\betaminus{2\\4}\zeta_5\nonumber\\&+\frac{2 i \pi^5\Bar{\tau}^5 \zeta_3}{14175}-\frac{i\pi^3\Bar{\tau}^3\zeta_5}{675}+\frac{2\zeta_3\zeta_5}{15}\,,
		\label{eq: betaeqvexample}
	\end{align}
	which is a modular form of weights $(0,-6)$.
	
	\subsection{Differential equation}
	\label{subsection: diffeqn}
	The iterated integrals $\betaeqvtau{j_1&\cdots&j_\ell\\k1&\cdots&k_\ell}$ satisfy a differential equation given by \cite{Brown:2017qwo,Dorigoni:2022npe,Dorigoni:2024oft}
	\begin{align}
		\pi\nabla_0\betaeqvtau{j_1&j_2&\cdots&j_\ell\\k_1&k_2&\cdots&k_\ell}&=-\frac{1}{4}\sum_{i=1}^\ell(k_i-j_i-2)\betaeqv{j_1&\cdots&j_i+1&\cdots&j_\ell\\k1&\cdots&k_i&\cdots&k_\ell}\nonumber\\&\quad+\frac{1}{4}\delta_{j_\ell,k_\ell-2}(2 i\tau_2)^{k_\ell}\GG_{k_\ell}(\tau)\betaeqvtau{j_1&j_2&\cdots&j_{\ell-1}\\k_1&k_2&\cdots&k_{\ell-1}}+\dots\,,
		\label{eq: betaeqvdiffeqn}
	\end{align}
	where the ellipsis refers to holomorphic cusp forms multiplying $\beta^{\mathrm{eqv}}$ of depth $\leq \ell-2$. As the objects we would like to express in terms of iterated integrals are MGFs, we already know from the construction of the tree-representations that we will not encounter iterated integrals of cusp forms. Hence they will not appear in any differential equation in the algorithm presented in section \ref{section: alg}, meaning we can ignore the terms appearing in the ellipsis.
	
	Equation \ref{eq: betaeqvdiffeqn} is a first order differential equation that can be solved within the space of $\beta^{\mathrm{eqv}}$ under specific conditions. These conditions follow from solving a linear algebra problem that we do not study in this work. As the EIEIs contain MGFs, this condition will always be satisfied in the examples of interest. The boundary conditions are in general provided by the fact that both sides of the equations have to represent modular objects. This fixes the integration constants in all cases except when the solution is modular invariant: as any constant is modular invariant, we can add a constant to the solution without changing the modular properties. However, adding a constant to a modular invariant $\beta^{\mathrm{eqv}}$ propagates to other $\beta^{\mathrm{eqv}}$'s through the differential equations (\ref{eq: betaeqvdiffeqn}). Therefore, the consistency of the system of differential equations (\ref{eq: betaeqvdiffeqn}) and modularity fixes most of the constants of the modular invariant $\beta^{\mathrm{eqv}}$'s. The remaining constants can be fixed by gauge-fixing the generating series of $\beta^{\mathrm{eqv}}$'s explained in \cite{Dorigoni:2022npe,Dorigoni:2024iyt,Dorigoni:2024oft}. Note that these are not the constants we alluded to in earlier sections, which still need to be fixed in the algorithm. This will be discussed in \ref{subsection: constants}.
	
	\section{From trees to equivariant iterated Eisenstein integrals}
	Having discussed the objects that we want to translate into one another in sections \ref{section: MGFs} and \ref{section: Iterint}, we will talk about the translation procedure in this section. The idea of the algorithm is to use the trees that we built in section \ref{subsection: trees} from the MGFs to integrate back up iteratively using the differential equation from section \ref{subsection: diffeqn}. This provides us with an iterated integral description of any convergent MGF free of holomorphic subgraphs, i.e. the algorithm provides us with a dictionary between lattice-sum and iterated integral representations of MGFs. 
	
	\label{section: alg}
	\subsection{Integrating back up along trees}
	\label{subsection: backup}
	The tree-representations of MGFs introduced in section \ref{subsection: trees} provide us with the instructions on how to write the lattice-sum MGF in terms of the iterated integrals $\beta^{\mathrm{eqv}}$. Furthermore, we know that the $\beta^{\mathrm{eqv}}$ satisfy a differential equation given in (\ref{eq: betaeqvdiffeqn}). Using this differential equation, we can start to integrate our way back up the tree, which will be explained step by step in this section.
	
	We start at the bottom of the tree, which consists of a family of holomorphic Eisenstein series with $\QQ$-coefficients. We can view each separate branch as the RHS of (\ref{eq: betaeqvdiffeqn}) and try to find the Left-Hand Side (LHS). We start with an ansatz that consists of all $\beta^{\mathrm{eqv}}$ such that acting with $\pi\nabla_0$ on any of them contains $\beta^{\mathrm{eqv}}$ on the RHS. We call such an object a candidate and denote it $\beta^\mathrm{eqv}_\mathrm{c}$ (for a single $\GG_k$ on the RHS this will always be just one candidate, but we will describe the general procedure where we could have many objects on the RHS here). Note that the set of candidates is severely restricted by the modular weight, the set of $k$-labels and depth of the $\beta^{\mathrm{eqv}}$ on the RHS as can be inferred from (\ref{eq: betaeqvdiffeqn}). In mathematical syntax, they are all the $\beta^{\mathrm{eqv}}_\mathrm{c}$ such that
	\begin{align}
		\pi\nabla_0\beta^{\mathrm{eqv}}_{\mathrm{c}}\cap  \mathrm{RHS}\neq \varnothing
		\,,
		\label{eq: candidates}
	\end{align}
	where in an abuse of language we represent the sets of $\beta^{\mathrm{eqv}}$ contained in $\pi\nabla_0\beta^{\mathrm{eqv}}_{\mathrm{c}}$ and RHS by the expressions themselves.
	
	Now it could be that besides creating the $\beta^\mathrm{eqv}$ we want, the candidates also produce certain $\beta^{\mathrm{eqv}}$ that are not contained in the RHS. These unwanted $\beta^{\mathrm{eqv}}$ should obviously not be there in the final expression, which we can realize by adding more candidates that gives rise to the same unwanted terms to cancel them. We will always look for all other possible candidates that could cancel the unwanted terms. Mathematically we look for the additional $\beta^{\mathrm{eqv}}_{\mathrm{c,add}}$ such that
	\begin{align}
		\pi\nabla_0\beta^{\mathrm{eqv}}_{\mathrm{c,add}}\cap\pi\nabla_0\beta^{\mathrm{eqv}}_{\mathrm{c}}\neq \varnothing\,,
		\label{eq: morecandidates}
	\end{align}
	where again we use the abuse of set-notation. This gives a new set of candidates that might be bigger than the old set. We keep doing this until the sets are equal, which is guaranteed to happen as the set of all possible candidates for given total degree $k$ and modular weights is finite. Now we have a proper ansatz for the LHS that consists of an arbitrary linear combination of all the candidate $\beta^\mathrm{eqv}$\footnote{There is another brute force way to do this. Given the degree, depth(s) and modular weight of the RHS, we can consider all $\beta^{\mathrm{eqv}}$ that give rise to the $\beta^{\mathrm{eqv}}$ of that specific degree, modular weight and those depths, which is generally bigger than the space of candidates through the other method. Then taking an arbitrary linear combination of them and acting with $\pi\nabla_0$, we can again form the linear system of equations and solve it. We are guaranteed to get the same solution by the linear independence of the iterated integrals.}. As all the distinct $\beta^\mathrm{eqv}$ are linearly independent \cite{matthes2017algebraic}, we can find the exact coefficients through a linear algebra problem where we act with $\pi\nabla_0$ on the ansatz and equate it with the original LHS. This gives a solvable linear set of equations such that we can find the coefficients. This gives the complete LHS that produces the RHS upon the action of $\pi\nabla_0$. 
	
	The process outlined above allowed us to integrate back up the tree from the bottom to the first level above the bottom. Now to get to the next level, we view the expression we found for the LHS as the RHS of equation (\ref{eq: betaeqvdiffeqn}) and try to find the LHS yet again. We keep doing this until we arrive at the top of the tree. However, three things can happen along the way that we need to consider. First, whenever we encounter a $\GG_k$ at the vertex which represents your new ``RHS'', we multiply it with this $\GG_k$ to get your actual new RHS. Second, whenever two branches combine in a single vertex, we add the corresponding new RHSs. Last, whenever we arrive at an orange vertex, we know that the original MGF was modular invariant. Therefore, as already discussed in section \ref{subsection: trees}, we might have an integration constant to fix. To fix such a constant, we need to match the constants of the cuspidal expansions of both the lattice-sum and iterated integral descriptions. These constants can be found for any MGF up to weight $|A|+|B|\leq 12$ through the \textsc{Mathematica} package \cite{Gerken:2020aju} and for specific examples of higher weight in \cite{Green:2008uj,Zerbini:2015rss}. The cuspidal expansions of the iterated integrals can be calculated from appendix \ref{appendix: LP}. 
	We show an explicit example of this in more detail in section \ref{subsection: constants}. 
	
	Once we have repeated this until we are back at the top of the tree, we have found the iterated integral expression of the MGF in terms of the $\beta^{\mathrm{eqv}}$, which is exactly what the goal was. To make all of this more concrete, let us look at examples. 
	
	\subsubsection{Examples}
	\label{subsubsection: examples}
	As we constructed the tree for $\cformp{2\,1\,1\\2\,1\,1}$ explicitly in section \ref{subsection: trees}, let us try to find its representation in terms of iterated integrals. We start at the bottom level where we find two separate branches, hence we have two separate RHSs of equation (\ref{eq: betaeqvdiffeqn}). Let us look at them one by one.
	\begin{align}
		\pi \nabla_0 \beta^{\mathrm{eqv}}_{\mathrm{c}}=672\tau_2^8\GG_8\,,\qquad \pi \nabla_0 \beta^{\mathrm{eqv}}_{\mathrm{c}}=-36\tau_2^4\GG_4\,.
		\label{eq: backup1stcand}
	\end{align}
	According to the differential equation in (\ref{eq: betaeqvdiffeqn}), the only candidate for the left equation is $\betaeqv{6\\8}$, while the only one for the right one is $\betaeqv{2\\4}$. Furthermore, they only produce the original RHSs, hence we found all candidates after one iteration. Now we form linear combinations (in this case just one object with a coefficient for both of them) and act with the differential operator to form the linear systems of equations
	\begin{align}
		\pi\nabla_0 \left( c_1 \betaeqv{6\\8}\right)&=64\tau_2^8 c_1 \GG_8=756\tau_2^8\GG_8\implies c_1=\frac{189}{16}\nonumber\,,\\\pi\nabla_0 \left( c_2 \betaeqv{2\\4}\right)&=4\tau_2^4 c_2 \GG_4=-36\tau_2^4\GG_4\implies c_2=-9\,.
		\label{eq: backup1st}
	\end{align}
	At this point we have moved one level up the tree, i.e. visually
	
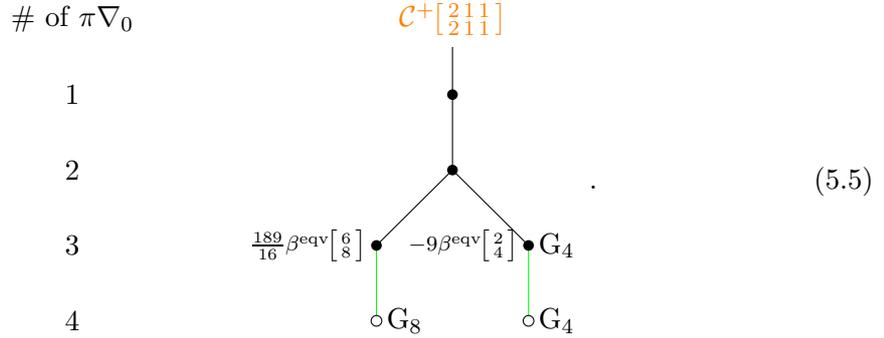
\begin{align}
		\begin{tikzpicture}[baseline=(zero.base)]
			\draw (0,0) to (0,-1);
			\draw (0,-1) to (0,-2);
			\draw (0,-2) to (-1,-3);
			\draw (0,-2) to (1,-3);
			\draw[color=green] (-1,-3) to (-1,-4);
			\draw[color=green] (1,-3) to (1,-4);
			\fill (0,-1) circle (2pt);
			\fill (0,-2) circle (2pt);
			\fill (-1,-3) circle (2pt);
			\fill (1,-3) circle (2pt);
			\node at (0,-9/4) (zero) {};
			\node[fill=white] at (0,0) {\textcolor{orange}{$\cformp{2\,1\,1\\2\,1\,1}$}};
			\node[right] at (1,-3) {$\GG_4$};
			\draw[fill=white] (-1,-4) circle[radius=2pt];
			\draw[fill=white] (1,-4) circle[radius=2pt];
			\node[right] at (1,-4) {$\GG_4$};
			\node[right] at (-1,-4) {$\GG_8$};
			\node[left] at (-1,-3) {\scriptsize{$\tfrac{189}{16}\betaeqv{6\\8}$}};
			\node[left] at (1,-3) {\scriptsize{$-9\betaeqv{2\\4}$}};
			\node[fill=white] at (-5,0) {$\#$ of $\pi\nabla_0$};
			\node[fill=white] at (-5,-1) {$1$};
			\node[fill=white] at (-5,-2) {$2$};
			\node[fill=white] at (-5,-3) {$3$};
			\node[fill=white] at (-5,-4) {$4$};
		\end{tikzpicture}\,.
		\label{eq: backup1sttree}
	\end{align}
	
	Now that we have arrived at level three, we see that for the right branch there is a $\GG_4$ attached to the vertex, hence according to the rules, we multiply to get $-9\betaeqv{2\\4}\tau_2^4\GG_4$. Again we repeat the procedure of finding candidates for these expressions, i.e. we want to find
	\begin{align}
		\pi\nabla_0\beta^{\mathrm{eqv}}_\mathrm{c}=\frac{189}{16}\betaeqv{6\\8}\,,\qquad \pi\nabla_0\beta^{\mathrm{eqv}}_\mathrm{c}=-9\betaeqv{2\\4}\tau_2^4\GG_4\,.
		\label{eq: backup2ndcand}
	\end{align}
	For the left equation this is again straightforward as there's only one final candidate, which is $\betaeqv{5\\8}$. For the right equation the same is true, and the only final candidate is $\betaeqv{2&2\\4&4}$. We form again the linear combinations for unknown coefficients, act with $\pi\nabla_0$ and solve for them. This moves us yet another level up.
	
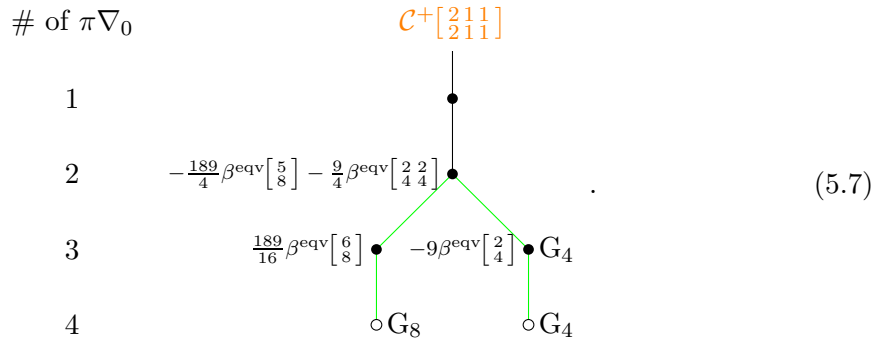
\begin{align}
		\begin{tikzpicture}[baseline=(zero.base)]
			\draw (0,0) to (0,-1);
			\draw (0,-1) to (0,-2);
			\draw[color=green] (0,-2) to (-1,-3);
			\draw[color=green] (0,-2) to (1,-3);
			\draw[color=green] (-1,-3) to (-1,-4);
			\draw[color=green] (1,-3) to (1,-4);
			\fill (0,-1) circle (2pt);
			\fill (0,-2) circle (2pt);
			\fill (-1,-3) circle (2pt);
			\fill (1,-3) circle (2pt);
			\node at (0,-9/4) (zero) {};
			\node[fill=white] at (0,0) {\textcolor{orange}{$\cformp{2\,1\,1\\2\,1\,1}$}};
			\node[right] at (1,-3) {$\GG_4$};
			\draw[fill=white] (-1,-4) circle[radius=2pt];
			\draw[fill=white] (1,-4) circle[radius=2pt];
			\node[right] at (1,-4) {$\GG_4$};
			\node[right] at (-1,-4) {$\GG_8$};
			\node[left] at (-1,-3) {\scriptsize{$\tfrac{189}{16}\betaeqv{6\\8}$}};
			\node[left] at (1,-3) {\scriptsize{$-9\betaeqv{2\\4}$}};
			\node[left] at (0,-2) {\scriptsize{$-\tfrac{189}{4}\betaeqv{5\\8}-\tfrac{9}{4}\betaeqv{2&2\\4&4}$}};
			\node[fill=white] at (-5,0) {$\#$ of $\pi\nabla_0$};
			\node[fill=white] at (-5,-1) {$1$};
			\node[fill=white] at (-5,-2) {$2$};
			\node[fill=white] at (-5,-3) {$3$};
			\node[fill=white] at (-5,-4) {$4$};
		\end{tikzpicture}\,.
		\label{eq: backup2nd}
	\end{align}
	As the separate branches now merged, we added the results at the vertex. Now we just have a single branch left over and thus a single expression. Finding the candidates is straightforward using the differential equation (\ref{eq: betaeqvdiffeqn}) again. We find 
	\begin{align}
		&\pi\nabla_0\left(c_3\betaeqv{4\\8}+c_4\betaeqv{2&1\\4&4}+c_5\betaeqv{1&2\\4&4}\right)\nonumber\\&=-\frac{c_3}{2}\betaeqv{5\\8}-\frac{c_4+c_5}{4}\betaeqv{2&2\\4&4}+4c_5\betaeqv{1\\4}\tau_2^4\GG_4=-\frac{189}{4}\betaeqv{5\\8}-\tfrac{9}{4}\betaeqv{2&2\\4&4}\nonumber\\&\implies c_3=\frac{189}{2},\quad c_4=9,\quad c_5=0\,.
		\label{eq: backup3rdcand}
	\end{align}
	We arrived at level 1 of the tree now
	\begin{align}
		\begin{tikzpicture}[baseline=(zero.base)]
			\draw (0,0) to (0,-1);
			\draw[color=green] (0,-1) to (0,-2);
			\draw[color=green] (0,-2) to (-1,-3);
			\draw[color=green] (0,-2) to (1,-3);
			\draw[color=green] (-1,-3) to (-1,-4);
			\draw[color=green] (1,-3) to (1,-4);
			\fill (0,-1) circle (2pt);
			\fill (0,-2) circle (2pt);
			\fill (-1,-3) circle (2pt);
			\fill (1,-3) circle (2pt);
			\node at (0,-9/4) (zero) {};
			\node[fill=white] at (0,0) {\textcolor{orange}{$\cformp{2\,1\,1\\2\,1\,1}$}};
			\node[right] at (1,-3) {$\GG_4$};
			\draw[fill=white] (-1,-4) circle[radius=2pt];
			\draw[fill=white] (1,-4) circle[radius=2pt];
			\node[right] at (1,-4) {$\GG_4$};
			\node[right] at (-1,-4) {$\GG_8$};
			\node[left] at (-1,-3) {\scriptsize{$\tfrac{189}{16}\betaeqv{6\\8}$}};
			\node[left] at (1,-3) {\scriptsize{$-9\betaeqv{2\\4}$}};
			\node[left] at (0,-2) {\scriptsize{$-\tfrac{189}{4}\betaeqv{5\\8}-\tfrac{9}{4}\betaeqv{2&2\\4&4}$}};
			\node[left] at (0,-1) {\scriptsize{$\tfrac{189}{2}\betaeqv{4\\8}+9\betaeqv{2&1\\4&4}$}};
			\node[fill=white] at (-5,0) {$\#$ of $\pi\nabla_0$};
			\node[fill=white] at (-5,-1) {$1$};
			\node[fill=white] at (-5,-2) {$2$};
			\node[fill=white] at (-5,-3) {$3$};
			\node[fill=white] at (-5,-4) {$4$};
		\end{tikzpicture}\,.
		\label{eq: backup3rdtree}
	\end{align}
	One final step remains. As it consists of yet again the same steps, we give the final answer
	\begin{align}
		\cformp{2\,1\,1\\2\,1\,1}&=-126\betaeqv{3\\8}-18\betaeqv{0&2\\4&4}\nonumber\\&=-126\betaplus{3\\8}-126\betaminus{3\\8}-18 \betaplus{0&2\\4&4}-18\betaminus{2&0\\4&4}-18\betaplus{2\\4}\betaminus{0\\4}\nonumber\\&\qquad-\frac{\pi i \Bar{\tau}\zeta_3}{30}-\frac{\pi\Bar{\tau}^2\zeta_3}{60\tau_2}+\frac{3\zeta_3}{4\pi^2\tau_2^2}\betaplus{2\\4}+12\zeta_3\betaminus{0\\4}-\frac{5\zeta_5}{12\pi\tau_2}-\frac{\zeta_3^2}{4\pi^2\tau_2^2}+\frac{9\zeta_7}{16\pi^3\tau_2^3}\,.
		\label{eq: c211beta}
	\end{align}
	Note that because $\cformp{2\,1\,1\\2\,1\,1}$ is a modular graph function, there was a constant to fix by comparing cuspidal expansions. In this case the constant turned out to be zero, but this does not have to be the case. We will consider such examples in more detail in the next section.
	
	A more nontrivial example would be that of $\cformtrip{1\,1\\1\,1}{1\,1\\1\,1}{1\,1\\1\,1}$. Its tree is
	\begin{align}
		\begin{tikzpicture}[baseline={([yshift=-0.5ex]current bounding box.center)}]
			\draw[] (29/4,1) to (29/4,0); 
			\draw[] (29/4,0) to (5,-1); 
			\draw[] (29/4,0) to (19/2,-1); 
			\draw[] (5,-1) to (2,-2);
			\draw[] (5,-1) to (8,-2); 
			\draw[] (5,-1) to (25/4,-2); 
			\draw[] (19/2,-1) to (19/2,-2); 
			\draw[] (2,-2) to (1/2,-3);
			\draw[] (2,-2) to (3,-3); 
			\draw[] (2,-2) to (4,-3); 
			\draw[] (25/4,-2) to (11/2,-3); 
			\draw[] (25/4,-2) to (7,-3); 
			\draw[] (19/2,-2) to (9,-3); 
			\draw[] (19/2,-2) to (10,-3);
			\draw[] (8,-2) to (8,-3); 
			\draw[] (1/2,-3) to (-1,-4);
			\draw[] (1/2,-3) to (0,-4);
			\draw[] (1/2,-3) to (1,-4);
			\draw[] (1/2,-3) to (2,-4);
			\draw[] (3,-3) to (3,-4);
			\draw[] (4,-3) to (4,-4);
			\draw[] (11/2,-3) to (5,-4);
			\draw[] (11/2,-3) to (6,-4);
			\draw[] (7,-3) to (7,-4);
			\draw[] (8,-3) to (8,-4); 
			\draw[] (9,-3) to (9,-4);
			\draw[] (10,-3) to (10,-4);
			\draw[] (-1,-4) to (-1,-5); 
			\draw[] (0,-4) to (0,-5);
			\draw[] (1,-4) to (1,-5);
			\draw[] (2,-4) to (2,-5);
			\draw[] (3,-4) to (3,-5);
			\draw[] (4,-4) to (4,-5);
			\draw[] (5,-4) to (5,-5);
			\draw[] (6,-4) to (6,-5);
			\draw[] (7,-4) to (7,-5);
			\draw[] (8,-4) to (8,-5);
			\draw[] (9,-4) to (9,-5);
			\draw[] (10,-4) to (10,-5);
			\fill (-1,-4) circle (2pt);
			\fill (0,-4) circle (2pt);
			\fill (1,-4) circle (2pt);
			\fill (2,-4) circle (2pt);
			\fill (3,-4) circle (2pt);
			\fill (4,-4) circle (2pt);
			\fill (5,-4) circle (2pt);
			\fill (6,-4) circle (2pt);
			\fill (7,-4) circle (2pt);
			\fill (8,-4) circle (2pt);
			\fill (9,-4) circle (2pt);
			\fill (10,-4) circle (2pt);
			\fill (1/2,-3) circle (2pt);
			\fill (3,-3) circle (2pt);
			\fill (4,-3) circle (2pt);
			\fill (11/2,-3) circle (2pt);
			\fill[color=orange] (7,-3) circle (2pt);
			\fill (8,-3) circle (2pt);
			\fill (9,-3) circle (2pt);
			\fill (10,-3) circle (2pt);
			\fill (2,-2) circle (2pt);
			\fill (25/4,-2) circle (2pt);
			\fill[color=orange] (8,-2) circle (2pt);
			\fill (19/2,-2) circle (2pt);
			\fill (5,-1) circle (2pt);
			\fill[color=orange] (19/2,-1) circle (2pt);
			\fill (29/4,0) circle (2pt);
			\draw[fill=white] (-1,-5) circle[radius=2pt];
			\draw[fill=white] (0,-5) circle[radius=2pt];
			\draw[fill=white] (1,-5) circle[radius=2pt];
			\draw[fill=white] (2,-5) circle[radius=2pt];
			\draw[fill=white] (3,-5) circle[radius=2pt];
			\draw[fill=white] (4,-5) circle[radius=2pt];
			\draw[fill=white] (5,-5) circle[radius=2pt];
			\draw[fill=white] (6,-5) circle[radius=2pt];
			\draw[fill=white] (7,-5) circle[radius=2pt];
			\draw[fill=white] (8,-5) circle[radius=2pt];
			\draw[fill=white] (9,-5) circle[radius=2pt];
			\draw[fill=white] (10,-5) circle[radius=2pt];
			\node[right] at (-1,-5) {$\GG_{12}$};
			\node[right] at (0,-5) {$\GG_8$}; 
			\node[right] at (1,-5) {$\GG_6$}; 
			\node[right] at (2,-5) {$\GG_4$}; 
			\node[right] at (3,-5) {$\GG_8$}; 
			\node[right] at (4,-5) {$\GG_6$}; 
			\node[right] at (5,-5) {$\GG_8$}; 
			\node[right] at (6,-5) {$\GG_4$}; 
			\node[right] at (7,-5) {$\GG_4$}; 
			\node[right] at (8,-5) {$\GG_6$}; 
			\node[right] at (9,-5) {$\GG_4$}; 
			\node[right] at (10,-5) {$\GG_4$}; 
			\node[right] at (0,-4) {$\GG_4$};
			\node[right] at (1,-4) {$\GG_6$};
			\node[right] at (2,-4) {$\GG_8$};
			\node[right] at (6,-4) {$\GG_4$};
			\node[right] at (9,-4) {$\GG_4$};
			\node[right] at (3,-3) {$\GG_4$};
			\node[right] at (4,-3) {$\GG_6$};
			\node[right] at (7,-3) {$\GG_4$};
			\node[right] at (10,-3) {$\GG_4$};
			\node[right] at (25/4,-2) {$\GG_4$};
			\node[right] at (8,-2) {$\GG_6$};
			\node[right] at (19/2,-1) {$\GG_4$};
			\node[fill=white] at (29/4,1) {\textcolor{orange}{$\cformtrip{1\,1\\1\,1}{1\,1\\1\,1}{1\,1\\1\,1}$}};    
			\node[fill=white] at (-2,1) {$\#$ of $\pi\nabla_0$};
			\node[fill=white] at (-2,0) {$1$};
			\node[fill=white] at (-2,-1) {$2$};
			\node[fill=white] at (-2,-2) {$3$};
			\node[fill=white] at (-2,-3) {$4$};
			\node[fill=white] at (-2,-4) {$5$};
			\node[fill=white] at (-2,-5) {$6$};
		\end{tikzpicture}\,.
		\label{eq: complicatedtree}
	\end{align}
	Performing all the steps in the algorithm we get
	\begin{align}
		\cformtrip{1\,1\\1\,1}{1\,1\\1\,1}{1\,1\\1\,1}&=-8712 \betaeqv{5\\12} - 1512 \betaeqv{2&2\\4&8} +7200 \betaeqv{2&2\\6&6} -24000 \betaeqv{3&1\\6&6}\nonumber\\&\qquad +21000 \betaeqv{4&0\\6&6} - 1512 \betaeqv{4&0\\8&4}-1296 \betaeqv{1&1&1\\4&4&4} +1296 \betaeqv{2&1&0\\4&4&4}\,.
		\label{eq: c222 example}
	\end{align}
	Even though in (\ref{eq: complicatedtree}) there appeared several modular graph functions, matching their cuspidal expansions did not introduce further constants. Further examples of modular graph functions in terms of iterated integrals can be found in the ancillary file of \cite{Claasen:2024ssh}.
	
	The depths and degrees of the iterated integrals involved are immediate from the tree representations; For every $\GG_{k_i}$ along a branch, the iterated integral increases by one unit of depth with lower entry $k_i$, which can easily be seen from the differential equation (\ref{eq: betaeqvdiffeqn}). These trees can therefore also be used to diagnose the drop in complexity for $n$-loop modular graph forms to depth $\leq n$ iterated integrals. The degree simply follows from the sum of the indices $k_i$ along any branch.
	
	\subsection{Dealing with constants}
	\label{subsection: constants}
	The fact that we are able to repeatedly act with a differential operator on a modular graph form and integrate back up to get its representation in terms of iterated integrals is remarkable when you consider the amount of integration constants that need to be fixed. Fortunately, many of these constants are fixed by the modularity of the iterated integrals. Even more so, writing the result of integration purely in terms of $\beta^{\mathrm{eqv}}$ automatically ensures the modularity of the integral. This upshot of using equivariant iterated integrals instead of any other type of iterated integrals of holomorphic Eisenstein series simplifies fixing these integration constants considerably. The only times when merely writing the integral in terms of equivariant iterated integrals is not enough to fix all the ambiguity, is whenever the integral describing the MGF is a modular function. This is because the constant part of the $\beta^{\mathrm{eqv}}$'s do not have to match with the constant parts of the MGFs, so that in this case we have an unknown integration constant that we need to fix. This is exactly why we colored certain vertices of the tree orange so that we keep track of whenever this happens.
	
	This phenomenon is best illustrated in the following example. Consider the modular graph functions $D_3:=\cformp{1\,1\,1\\1\,1\,1}$ and $\EE_3$. Their trees are identical
	\begin{align}
		\begin{tikzpicture}[baseline={([yshift=-0.5ex]current bounding box.center)}]
			\draw[] (1,1) to (1,0); 
			\draw[] (1,0) to (1,-1);
			\draw[] (1,-1) to (1,-2);
			\draw[] (5,1) to (5,0);
			\draw[] (5,0) to (5,-1);
			\draw[] (5,-1) to (5,-2);
			\fill (1,0) circle (2pt);
			\fill (1,-1) circle (2pt);
			\fill (5,0) circle (2pt);
			\fill (5,-1) circle (2pt);
			\draw[fill=white] (1,-2) circle[radius=2pt];
			\draw[fill=white] (5,-2) circle[radius=2pt];
			\node[right] at (1,-2) {$\GG_6$}; 
			\node[right] at (5,-2) {$\GG_6$};
			\node[fill=white] at (1,1) {\textcolor{orange}{$D_3$}};    
			\node[fill=white] at (-2,1) {$\#$ of $\pi\nabla_0$};
			\node[fill=white] at (-2,0) {$1$};
			\node[fill=white] at (-2,-1) {$2$};
			\node[fill=white] at (-2,-2) {$3$};
			\node[fill=white] at (5,1) {\textcolor{orange}{$\EE_3$}};   
		\end{tikzpicture}\,.
		\label{eq: tree}
	\end{align}
	Naively integrating back up we get 
	\begin{align}
		D_3\overset{?}{=}-30\betaeqv{2\\6},\qquad \EE_3\overset{?}{=}-30\betaeqv{2\\6}\,.
	\end{align}
	The first equation is actually false, while the second equation is true. This is exactly due to the fact that both $D_3$ and $\EE_3$ are modular graph functions (i.e. they are modular invariant), which means that this last integration step introduces a possible constant we can add to the solution. So actually we have
	\begin{align}
		D_3=-30\betaeqv{2\\6}+c_1,\qquad \EE_3=-30\betaeqv{2\\6}+c_2\,.
	\end{align}
	These constants can be fixed by comparing the constants in the cuspidal expansions of both the lattice sum and the iterated integrals, i.e. we want the constant of the Laurent polynomial of them to agree with the constants of the $\beta^{\mathrm{eqv}}$ at the cusp. Matching them, we can fix the constants. We find in this case $c_1=\zeta_3$ and $c_2=0$.
	
	For orange vertices not accompanied by a factor of $\GG_k$, we do not have to add any constant as the integral of just a constant cannot be written within the space of $\beta^{\text{eqv}}$'s. A simple example of an orange vertex accompanied by a factor of $\GG_k$ is in the four-loop banana graph $D_5=\cformp{1\,1\,1\,1\,1\\1\,1\,1\,1\,1}$. Its tree is given by
	
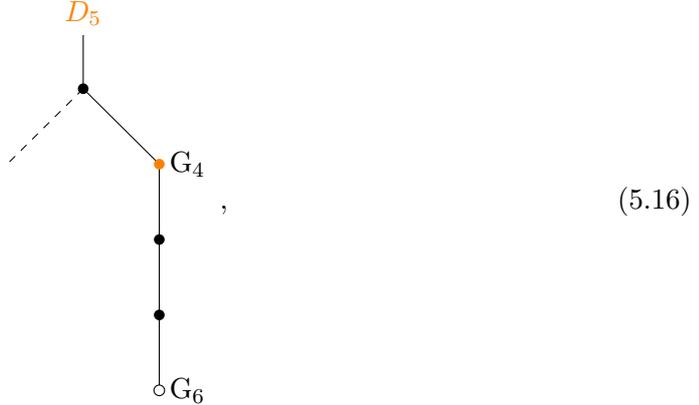
\begin{align}
		\begin{tikzpicture}[baseline={([yshift=-0.5ex]current bounding box.center)}]
			\draw[] (1,1) to (1,0); 
			\draw[] (1,0) to (2,-1);
			\draw[] (2,-1) to (2,-2);
			\draw[] (2,-2) to (2,-3);
			\draw[] (2,-3) to (2,-4);
			\draw[dashed] (1,0) to (0,-1);
			\fill (1,0) circle (2pt);
			\fill[color=orange] (2,-1) circle (2pt);
			\fill (2,-2) circle (2pt);
			\fill (2,-3) circle (2pt);
			\draw[fill=white] (2,-4) circle[radius=2pt];
			\node[right] at (2,-4) {$\GG_6$}; 
			\node[right] at (2,-1) {$\GG_4$}; 
			\node[fill=white] at (1,1) {\textcolor{orange}{$D_5$}};    
		\end{tikzpicture}\,,
		\label{eq: treed5}
	\end{align}
	where we omit the rest of the graph. The point is that the modular graph form that multiplies the $\GG_4$ is a modular graph function. In fact it is $D_3$, while the tree itself would suggest it could also be $\EE_3$. This introduces a term $\zeta_3 \GG_4$ that we need to add at that vertex, which propagates further upwards through the tree in the algorithm. In fact, we could have guessed the svMZV that enters this constant from uniform transcendentality of the Laurent polynomial of $D_5$ as follows. The transcendental weight of the Laurent polynomial of an MGF of weights $(a,b)$ at the top of a tree is $\tfrac{1}{2}(a+b)$. Each $\pi\nabla$ adds one unit of transcendental weight, while each $\GG_k$ you encounter removes $k$ units. Once you arrive at the orange vertex the transcendental weight of the svMZV must therefore equal $\tfrac{1}{2}(a+b)+N-\sum_{i\in I} k_i$ where $I$ is the index set of all $\GG_{k_i}$ encountered along the way and $N$ is the number of edges along the branch from the top up to the orange vertex. In the example above, the transcendental weight of the svMZV is $5+2-4=3$. The only svMZV at this weight is $\zeta^{\rm sv}_3=2\zeta_3$. 
	
	\subsection{Integrand of one-loop four-graviton superstring amplitude at order \texorpdfstring{$\alpha^{\prime 8}$}{TEXT}}
	\label{subsection: w8}
	The first application of the algorithm and the \textsc{Mathematica} package provided with this work appeared in \cite{Claasen:2024ssh}, where the integrand of the four-graviton Type II superstring amplitude was written in terms of iterated integrals up to seventh order in the inverse string tension $\alpha^{\prime 7}$. Up to that order, only $\beta^{\mathrm{eqv}}$ up to depth 3 appear. Here, we study the first instance involving $\beta^{\mathrm{eqv}}$ of depth four, which arises in the $\alpha^{\prime 8}$ term of the integrand of the low-energy one-loop four-graviton amplitude. The MGF expansion of the integrand can be derived from the Taylor expansion of the Koba-Nielsen factor \cite{Green:1999pv,DHoker:2015gmr}. In this expansion, $\alpha^{\prime}$ is absorbed in the Mandelstam invariants $s_{ij} = -\alpha^{\prime}(k_i+k_j)^2/4$, where $k_i$ and $k_j$ are the momenta of the gravitons. The genus expansion of the amplitude can be written as \cite{Green:2008uj,DHoker:2015gmr}
	\begin{equation}
		{A}= \kappa_{10}^2 \mathcal{R}^4\sum_{h=0}^\infty g_s^{2h-2} {A}^{(h)}(s_{ij}) \, ,
		\label{eq: fullamplitude}
	\end{equation}
	where $\kappa_{10}$ is the ten-dimensional gravitational constant. The factor $\mathcal{R}^4$ represents a contraction of four linearized Riemann tensors. $g_s$ is the string coupling and $h$ is the genus of the worldsheet. $A^{(h)}(s_{ij})$ are the genus-$h$ coefficient functions that depend on the Mandelstam variables. At genus one, we can write \cite{Green:1981yb}
	\begin{equation}
		{A}^{(1)}(s_{ij})=2\pi\int_{\mathcal{M}}\frac{d^2\tau}{\tau_2^2}\mathcal{I}(s_{ij}|\tau)\, ,
		\label{eq: g1amplitude}
	\end{equation}
	where $\mathcal{M}$ is fundamental domain for $\mathrm{SL}(2,\mathbb{Z})$ and $\mathcal{I}$ is the integrand. Expanding the integrand, we find the integrand $ \mathcal{I}$ at eighth order in $\alpha^{\prime}$ the following MGF expansion
	\begin{align}
		\mathcal{I}\bigg|_{\sigma_2\sigma_3^2}&(s_{ij}|\tau)=\tfrac{5}{324} \cformp{1\,1\\1\,1}\cformp{1\,1\,1\\1\,1\,1}^2 
		- \tfrac{1}{108} \cformp{1\,1\\1\,1}^2\cformp{1\,1\,1\,1\\1\,1\,1\,1} 
		- \tfrac{1}{1296} \cformp{1\,1\,1\,1\\1\,1\,1\,1}^2 \nonumber\\&
		+ \tfrac{11}{3240} \cformp{1\,1\,1\\1\,1\,1}\cformp{1\,1\,1\,1\,1\\1\,1\,1\,1\,1} 
		+ \tfrac{1}{3240} \cformp{1\,1\\1\,1}\cformp{1\,1\,1\,1\,1\,1\\1\,1\,1\,1\,1\,1} 
		+ \tfrac{1}{45360} \cformp{1\,1\,1\,1\\1\,1\,1\,1}\cformtrip{1\\1}{1\\1}{1\\1} \nonumber\\&
		- \tfrac{1}{36} \cformp{1\,1\,1\\1\,1\,1}\cformtrip{1\,1\\1\,1}{1\,1\\1\,1}{1\\1} 
		+ \tfrac{1}{36} \cformp{1\,1\\1\,1}\cformtrip{1\,1\\1\,1}{1\,1\\1\,1}{1\,1\\1\,1} 
		+ \tfrac{1}{27} \cformp{1\,1\,1\\1\,1\,1}\cformtrip{1\,1\,1\\1\,1\,1}{1\\1}{1\\1} \nonumber\\&
		- \tfrac{1}{18} \cformp{1\,1\\1\,1}\cformtrip{1\,1\,1\\1\,1\,1}{1\,1\\1\,1}{1\\1} 
		+ \tfrac{1}{36} \cformp{1\,1\\1\,1}\cformtrip{1\,1\,1\,1\\1\,1\,1\,1}{1\\1}{1\\1} 
		- \tfrac{1}{324} \cformtrip{1\,1\,1\\1\,1\,1}{1\,1\\1\,1}{1\,1\,1\\1\,1\,1} \nonumber\\&
		+ \tfrac{1}{648} \cformtrip{1\,1\,1\,1\\1\,1\,1\,1}{1\\1}{1\,1\,1\\1\,1\,1} 
		+ \tfrac{1}{216} \cformtrip{1\,1\,1\,1\\1\,1\,1\,1}{1\,1\\1\,1}{1\,1\\1\,1} 
		- \tfrac{1}{216} \cformtrip{1\,1\,1\,1\,1\\1\,1\,1\,1\,1}{1\\1}{1\,1\\1\,1} \nonumber\\&
		+ \tfrac{1}{648} \cformtrip{1\,1\,1\,1\,1\,1\\1\,1\,1\,1\,1\,1}{1\\1}{1\\1} 
		+ \tfrac{1}{36} \cformboxp{1\\1}{1\\1}{1\,1\\1\,1}{1\,1\,1\,1\\1\,1\,1\,1} 
		- \tfrac{1}{72} \cformboxp{1\,1\\1\,1}{1\,1\\1\,1}{1\,1\\1\,1}{1\,1\\1\,1} 
		\nonumber\\&
		- \tfrac{1}{36} \cformboxp{1\,1\,1\\1\,1\,1}{1\\1}{1\\1}{1\,1\,1\\1\,1\,1} 
		+ \tfrac{1}{36} \cformboxp{1\,1\,1\\1\,1\,1}{1\\1}{1\,1\\1\,1}{1\,1\\1\,1} 
		- \tfrac{1}{540} \cformboxp{1\,1\,1\,1\,1\\1\,1\,1\,1\,1}{1\\1}{1\\1}{1\\1} \nonumber\\&
		+ \tfrac{1}{108} \cformkitep{1\\1}{1\\1}{1\\1}{1\\1}{1\,1\,1\,1\\1\,1\,1\,1} 
		- \tfrac{1}{27} \cformkitep{1\,1\\1\,1}{1\\1}{1\\1}{1\\1}{1\,1\,1\\1\,1\,1} 
		+ \tfrac{1}{36} \cformkitep{1\,1\\1\,1}{1\\1}{1\,1\\1\,1}{1\\1}{1\,1\\1\,1} \nonumber\\&
		- \tfrac{1}{36} \cformkitep{1\,1\\1\,1}{1\,1\\1\,1}{1\\1}{1\\1}{1\,1\\1\,1} 
		+ \tfrac{1}{36} \cformkitep{1\,1\\1\,1}{1\,1\\1\,1}{1\,1\\1\,1}{1\\1}{1\\1} 
		+ \tfrac{2}{27} \cformkitep{1\,1\,1\\1\,1\,1}{1\\1}{1\\1}{1\\1}{1\,1\\1\,1} \nonumber\\&
		- \tfrac{2}{27} \cformkitep{1\,1\,1\\1\,1\,1}{1\\1}{1\,1\\1\,1}{1\\1}{1\\1} 
		+ \tfrac{1}{54} \cformkitep{1\,1\,1\\1\,1\,1}{1\,1\\1\,1}{1\\1}{1\\1}{1\\1} 
		- \tfrac{5}{108} \cformkitep{1\,1\,1\,1\\1\,1\,1\,1}{1\\1}{1\\1}{1\\1}{1\\1} \nonumber\\&
		+ \tfrac{1}{36} \cformtetp{1\\1}{1\,1\\1\,1}{1\\1}{1\\1}{1\,1\\1\,1}{1\\1} 
		- \tfrac{1}{18} \cformtetp{1\,1\\1\,1}{1\\1}{1\,1\\1\,1}{1\\1}{1\\1}{1\\1} 
		+ \tfrac{1}{27} \cformtetp{1\,1\,1\\1\,1\,1}{1\\1}{1\\1}{1\\1}{1\\1}{1\\1} \, ,
	\end{align}
	and
	\begin{align}
		\mathcal{I}\bigg|_{\sigma_2^4}&(s_{ij}|\tau) =- \tfrac{1}{288} \cformp{1\,1\\1\,1}\cformp{1\,1\,1\\1\,1\,1}^2 
		+ \tfrac{1}{128} \cformp{1\,1\\1\,1}^2 \cformp{1\,1\,1\,1\\1\,1\,1\,1}
		+ \tfrac{1}{1536} \cformp{1\,1\,1\,1\\1\,1\,1\,1}^2 \nonumber\\&
		- \tfrac{1}{2880} \cformp{1\,1\,1\\1\,1\,1}\cformp{1\,1\,1\,1\,1\\1\,1\,1\,1\,1} 
		+ \tfrac{1}{1920} \cformp{1\,1\\1\,1}\cformp{1\,1\,1\,1\,1\,1\\1\,1\,1\,1\,1\,1} 
		+ \tfrac{1}{161280} \cformp{1\,1\,1\,1\,1\,1\,1\,1\\1\,1\,1\,1\,1\,1\,1\,1} \nonumber\\&
		- \tfrac{1}{192} \cformboxp{1\\1}{1\\1}{1\,1\\1\,1}{1\,1\,1\,1\\1\,1\,1\,1} 
		+ \tfrac{1}{256} \cformboxp{1\,1\\1\,1}{1\,1\\1\,1}{1\,1\\1\,1}{1\,1\\1\,1} 
		+ \tfrac{1}{96} \cformboxp{1\,1\,1\\1\,1\,1}{1\\1}{1\\1}{1\,1\,1\\1\,1\,1} \nonumber\\&
		- \tfrac{1}{96} \cformboxp{1\,1\,1\\1\,1\,1}{1\\1}{1\,1\\1\,1}{1\,1\\1\,1} 
		+ \tfrac{1}{480} \cformboxp{1\,1\,1\,1\,1\\1\,1\,1\,1\,1}{1\\1}{1\\1}{1\\1} \, ,
	\end{align}
	where $\sigma_k = s_{12}^k+s_{13}^k+s_{23}^k$ and we have imposed the on-shell condition and the conservation of external momenta. The notations of the topologies beyond equations (\ref{eq: dihedralC}) and (\ref{eq: trihedralC}) are given by \cite{Gerken:2020aju}
	\begin{align}
		\cformboxl{A_{1}\\B_{1}}{A_{2}\\B_{2}}{A_{3}\\B_{3}}{A_{4}\\B_{4}}
		=
		\begin{tikzpicture}[baseline=(zero.base)]
			\node (1) at (0,0)  [circle]{$1$};
			\node (2) at (0,3)  [circle]{$2$};
			\node (3) at (3,3)  [circle]{$3$};
			\node (4) at (3,0)  [circle]{$4$};
			\draw[->-] (1) to[bend left=15] (2);
			\draw[->-] (1) to[bend right=15] (2);
			\draw[->-] (1) to
			node(zero)[fill=white]{$\sbmatrix{A_{1}\\B_{1}}$}(2);
			\draw[->-] (2) to[bend left=15] (3);
			\draw[->-] (2) to[bend right=15] (3);
			\draw[->-] (2) to
			node[fill=white]{$\sbmatrix{A_{2}\\B_{2}}$}(3);
			\draw[->-] (3) to[bend left=15] (4);
			\draw[->-] (3) to[bend right=15] (4);
			\draw[->-] (3) to
			node[fill=white]{$\sbmatrix{A_{3}\\B_{3}}$}(4);
			\draw[->-] (4) to[bend left=15] (1);
			\draw[->-] (4) to[bend right=15] (1);
			\draw[->-] (4)
			to node[fill=white]{$\sbmatrix{A_{4}\\B_{4}}$}(1);
		\end{tikzpicture}\,,
		\label{eq: tetrahedralC}
	\end{align}
	called the box topology. Adding a diagonal block gives the so-called kite topology
	\begin{align}
		\cformkitel{A_{1}\\B_{1}}{A_{2}\\B_{2}}{A_{3}\\B_{3}}{A_{4}\\B_{4}}
		{A_{5}\\B_{5}}=
		\begin{tikzpicture}[baseline=(zero.base)]
			\node (1) at (0,0)  [circle]{$1$};
			\node (2) at (0,3.2)  [circle]{$2$};
			\node (3) at (3.2,3.2)  [circle]{$3$};
			\node (4) at (3.2,0)  [circle]{$4$};
			\draw[->-] (1) to[bend left=15] (2);
			\draw[->-] (1) to[bend right=15] (2);
			\draw[->-] (1) to
			node(zero)[fill=white]{$\sbmatrix{A_{1}\\B_{1}}$}(2);
			\draw[->-] (2) to[bend left=15] (3);
			\draw[->-] (2) to[bend right=15] (3);
			\draw[->-] (2) to
			node[fill=white]{$\sbmatrix{A_{2}\\B_{2}}$}(3);
			\draw[->-] (1) to[bend left=15] (4);
			\draw[->-] (1) to[bend right=15] (4);
			\draw[->-] (1) to
			node[fill=white]{$\sbmatrix{A_{3}\\B_{3}}$}(4);
			\draw[->-] (4) to[bend left=15] (3);
			\draw[->-] (4) to[bend right=15] (3);
			\draw[->-] (4) to
			node[fill=white]{$\sbmatrix{A_{4}\\B_{4}}$}(3);
			\draw[->-] (1) to[bend left=10] (3);
			\draw[->-] (1) to[bend right=10] (3);
			\draw[->-] (1) to
			node[fill=white]{$\sbmatrix{A_{5}\\B_{5}}$}(3);
		\end{tikzpicture}\,.
	\end{align}
	Adding one last diagonal block gives the full tetrahedral topology (also known as the Mercedes graph)
	\begin{align}
		\cformtet{A_{1}\\B_{1}}{A_{2}\\B_{2}}{A_{3}\\B_{3}}{A_{4}\\B_{4}}
		{A_{5}\\B_{5}}{A_{6}\\B_{6}}
		=
		\begin{tikzpicture}[baseline=(4.base)]
			\node (1) at (0,0)  [circle]{$1$};
			\node (2) at (60:5.5) [circle]{$2$};
			\node (3) at (5.5,0) [circle]{$3$};
			\draw[opacity=0,name path = ray1](0,0)--(30:5);
			\draw[opacity=0,name path = vert](2)--(2 |- 0,0);
			\node[name intersections={of=ray1 and vert}] (4) at (intersection-1) [circle]{$4$};
			\draw[->-] (4) to[bend left=10] (2);
			\draw[->-] (4) to[bend right=10] (2);
			\draw[->-] (4) to
			node[fill=white]{$\sbmatrix{A_{2}\\B_{2}}$}(2);
			\draw[->-] (1) to[bend left=10] (4);
			\draw[->-] (1) to[bend right=10] (4);
			\draw[->-] (1) to
			node[fill=white]{$\sbmatrix{A_{6}\\B_{6}}$}(4);
			\draw[->-] (1) to[bend left=7] (2);
			\draw[->-] (1) to[bend right=7] (2);
			\draw[->-] (1) to
			node[fill=white]{$\sbmatrix{A_{1}\\B_{1}}$}(2);
			\draw[->-] (4) to[bend left=10] (3);
			\draw[->-] (4) to[bend right=10] (3);
			\draw[->-] (4) to
			node[fill=white]{$\sbmatrix{A_{4}\\B_{4}}$}(3);
			\draw[->-] (3) to[bend left=7] (2);
			\draw[->-] (3) to[bend right=7] (2);
			\draw[->-] (3) to
			node[fill=white]{$\sbmatrix{A_{3}\\B_{3}}$}(2);
			\draw[->-] (3) to[bend left=7] (1);
			\draw[->-] (3) to[bend right=7] (1);
			\draw[->-] (3) to
			node[fill=white]{$\sbmatrix{A_{5}\\B_{5}}$}(1);
		\end{tikzpicture}\,.
		\label{eq: mercedesC}
	\end{align}
	\indent Using the package \textsc{MGFtoBeqv}, we get the following expansion in terms of $\beta^{\mathrm{eqv}}$ up to depth 4
	\begin{align}
		\mathcal{I}\bigg|_{\sigma_2\sigma_3^2}&(s_{ij}|\tau) = 801720 \betaeqv{7 \\ 16} - 25432 \betaeqv{1&5 \\ 4&12} + \frac{236104}{3} \betaeqv{2&4 \\ 4&12} + 18900 \betaeqv{2&4 \\ 6&10} \notag\\
		&+ 52560 \betaeqv{3&3 \\ 6&10} + 178520 \betaeqv{4&2 \\ 6&10} + 21952 \betaeqv{4&2 \\ 8&8} + 18900 \betaeqv{4&2 \\ 10&6} \notag\\
		&+ 228536 \betaeqv{5&1 \\ 8&8} + 52560 \betaeqv{5&1 \\ 10&6} - 25432 \betaeqv{5&1 \\ 12&4} + 81928 \betaeqv{6&0 \\ 8&8} \notag\\
		&+ 178520 \betaeqv{6&0 \\ 10&6} + \frac{236104}{3} \betaeqv{6&0 \\ 12&4} - 2016 \betaeqv{1&2&2 \\ 4&4&8} - 15000 \betaeqv{1&2&2 \\ 4&6&6} \notag\\
		&+ 11200 \betaeqv{1&3&1 \\ 4&6&6} - 10800 \betaeqv{1&4&0 \\ 4&6&6} - 2016 \betaeqv{1&4&0 \\ 4&8&4} + 2688 \betaeqv{2&1&2 \\ 4&4&8} \notag\\
		&+ 14000 \betaeqv{2&1&2 \\ 4&6&6} - 15000 \betaeqv{2&1&2 \\ 6&4&6} + 4200 \betaeqv{2&2&1 \\ 4&4&8} + 3600 \betaeqv{2&2&1 \\ 4&6&6} \notag\\
		&- 2016 \betaeqv{2&2&1 \\ 4&8&4} + 14000 \betaeqv{2&2&1 \\ 6&4&6} - 15000 \betaeqv{2&2&1 \\ 6&6&4} + \frac{41200}{3} \betaeqv{2&3&0 \\ 4&6&6} \notag\\
		&+ 7728 \betaeqv{2&3&0 \\ 4&8&4} + 14000 \betaeqv{2&3&0 \\ 6&6&4} + 14000 \betaeqv{3&0&2 \\ 6&4&6} + \frac{56000}{3} \betaeqv{3&1&1 \\ 6&4&6} \notag\\
		&+ 11200 \betaeqv{3&1&1 \\ 6&6&4} + \frac{400}{3} \betaeqv{3&2&0 \\ 6&4&6} + 3600 \betaeqv{3&2&0 \\ 6&6&4} + \frac{400}{3} \betaeqv{4&0&1 \\ 6&4&6} \notag\\
		&- 10800 \betaeqv{4&0&1 \\ 6&6&4} - 2016 \betaeqv{4&0&1 \\ 8&4&4} + 26800 \betaeqv{4&1&0 \\ 6&4&6} + \frac{41200}{3} \betaeqv{4&1&0 \\ 6&6&4} \notag\\
		&+ 2688 \betaeqv{4&1&0 \\ 8&4&4} + 4200 \betaeqv{5&0&0 \\ 8&4&4} - 432 \betaeqv{1&2&1&0 \\ 4&4&4&4} - 432 \betaeqv{2&1&0&1 \\ 4&4&4&4} \notag\\
		&+ 144 \betaeqv{2&1&1&0 \\ 4&4&4&4} + 504 \betaeqv{2&2&0&0 \\ 4&4&4&4} - 126 \betaeqv{4 \\ 10} \zeta_3 + 100 \betaeqv{1&2 \\ 4&6} \zeta_3 \notag\\
		&- \frac{280}{3} \betaeqv{2&1 \\ 4&6} \zeta_3 + 100 \betaeqv{2&1 \\ 6&4} \zeta_3 - \frac{280}{3} \betaeqv{3&0 \\ 6&4} \zeta_3 - \frac{1}{3} \betaeqv{1 \\ 4} \zeta_3^2 \notag\\
		&- \frac{145}{18} \betaeqv{2 \\ 6} \zeta_5 + c_1 \, ,
		\label{sigma2sigma3}
	\end{align}
	and
	\begin{align}
		\mathcal{I}\bigg|_{\sigma_2^4}&(s_{ij}|\tau)=-165600 \betaeqv{7 \\ 16} + 8844 \betaeqv{1&5 \\ 4&12} - 22110 \betaeqv{2&4 \\ 4&12} - 16920 \betaeqv{3&3 \\ 6&10} \notag\\
		&+ 3136 \betaeqv{3&3 \\ 8&8} - 33030 \betaeqv{4&2 \\ 6&10} - 14112 \betaeqv{4&2 \\ 8&8} - 35280 \betaeqv{5&1 \\ 8&8} \notag\\
		&- 16920 \betaeqv{5&1 \\ 10&6} + 8844 \betaeqv{5&1 \\ 12&4} - 18424 \betaeqv{6&0 \\ 8&8} - 33030 \betaeqv{6&0 \\ 10&6} \notag\\
		&- 22110 \betaeqv{6&0 \\ 12&4} - 2016 \betaeqv{1&1&3 \\ 4&4&8} + 3024 \betaeqv{1&2&2 \\ 4&4&8} + 2400 \betaeqv{1&3&1 \\ 4&6&6} \notag\\
		&- 2016 \betaeqv{1&3&1 \\ 4&8&4} + 1200 \betaeqv{1&4&0 \\ 4&6&6} + 3024 \betaeqv{1&4&0 \\ 4&8&4} + 1008 \betaeqv{2&0&3 \\ 4&4&8} \notag\\
		&- 3024 \betaeqv{2&2&1 \\ 4&4&8} - 3600 \betaeqv{2&2&1 \\ 4&6&6} + 3024 \betaeqv{2&2&1 \\ 4&8&4} - 3600 \betaeqv{2&3&0 \\ 4&6&6} \notag\\
		&- 5040 \betaeqv{2&3&0 \\ 4&8&4} - 2400 \betaeqv{3&1&1 \\ 6&4&6} + 2400 \betaeqv{3&1&1 \\ 6&6&4} - 2016 \betaeqv{3&1&1 \\ 8&4&4} \notag\\
		&- 1800 \betaeqv{3&2&0 \\ 6&4&6} - 3600 \betaeqv{3&2&0 \\ 6&6&4} + 1008 \betaeqv{3&2&0 \\ 8&4&4} - 1800 \betaeqv{4&0&1 \\ 6&4&6} \notag\\
		&+ 1200 \betaeqv{4&0&1 \\ 6&6&4} + 3024 \betaeqv{4&0&1 \\ 8&4&4} - 5100 \betaeqv{4&1&0 \\ 6&4&6} - 3600 \betaeqv{4&1&0 \\ 6&6&4} \notag\\
		&- 3024 \betaeqv{5&0&0 \\ 8&4&4} + 1296 \betaeqv{1&1&1&1 \\ 4&4&4&4} - 648 \betaeqv{1&1&2&0 \\ 4&4&4&4} - 648 \betaeqv{1&2&0&1 \\ 4&4&4&4} \notag\\
		&- 648 \betaeqv{2&0&1&1 \\ 4&4&4&4} + 324 \betaeqv{2&0&2&0 \\ 4&4&4&4} + 324 \betaeqv{2&2&0&0 \\ 4&4&4&4} + c_2 \, ,
		\label{sigma2}
	\end{align}
	where $c_1$ and $c_2$ are integration constants, which we know by transcendentality arguments to be proportional to $\zeta_3\zeta_5$. These constants form a crucial part of the analytic part of the amplitude when integrating the expressions over the fundamental domain. As discussed in \cite{Claasen:2024ssh}, $c_1$ and $c_2$ precisely determine the eighth order of the analytic part of the amplitude. Even though we cannot get individual constants in the Laurent polynomials of MGFs at this order, we can infer the combination of these constants that contribute to the amplitude through divergences of the one-loop matrix elements of the field theory limit \cite{Edison:2021ebi}. All that remains to determine $c_1$ and $c_2$ is the matching of the constants of the Laurent polynomials of $\beta^{\mathrm{eqv}}$ expansion with the constants from the UV-divergences. The combinations of constants of MGFs are given through the UV-divergences up to the ninth order of $\alpha^{\prime}$ by\footnote{We thank Alex Edison for providing us with the divergences of the supergravity one-loop matrix elements.} 
	\begin{align}
		\mathcal{I}_0 &= \frac{1}{3}\zeta_3\sigma_3+\frac{29}{135}\zeta_5\sigma_2\sigma_3+\frac{1}{30}\zeta_3^2\sigma_2^3+\frac{7}{135}\zeta_3^2\sigma_3^2+\frac{89}{2520}\zeta_7\sigma_2^2\sigma_3+\frac{286}{2835}\zeta_3\zeta_5 \sigma_2\sigma_3^2+ \frac{11}{420}\zeta_3\zeta_5 \sigma_2^4 \notag\\ & \quad+\left(\frac{43}{8505}\zeta_3^3+\frac{271}{17010}\zeta_9\right)\sigma_3^3 +\left(\frac{83}{5670}\zeta_3^3+\frac{1423}{22680}\zeta_9\right)\sigma_2^3\sigma_3 + \mathcal{O}(s_{ij}^{10})
	\end{align}
	The calculation of the Laurent polynomials of the $\beta^{\mathrm{eqv}}$ expansions is explained in appendix \ref{appendix: LP} and \ref{appendix: csv}. Matching the zero modes of the integrand, denoted $\mathcal{I}_0$, gives the equations
	\begin{align}
		\mathcal{I}_0\bigg|_{\sigma_2\sigma_3^2} &= \frac{286}{2835}\zeta_3\zeta_5 = c_1+\frac{154781}{2639952}\zeta_3\zeta_5 \, , \\
		\mathcal{I}_0\bigg|_{\sigma_2^4} &= \frac{11}{420}\zeta_3\zeta_5 = c_2+\frac{55229}{2933280}\zeta_3\zeta_5 \, ,
	\end{align}
	which yields $c_1 = \frac{79673}{97\times 3^3 \times 6!} \zeta_3\zeta_5$ and $c_2 = \frac{617}{97\times 6^2\times 4!} \zeta_3\zeta_5$. Following \cite{Green:2000,Green:2008uj}, we split the amplitude $A^{(1)}$ into an analytic and a non-analytic part\footnote{There is a slight abuse of language in using analytic versus non-analytic as the non-analytic part contains polynomials of the Mandelstam invariants which are analytic. The analytic part contains all svMZVs with maximal transcendental weights at each order in $\alpha^{\prime}$ \cite{DHoker:2019blr,Claasen:2024ssh}}. We conclude this section by integrating the constants over the fundamental domain. The integral of a constant is proportional to
	\begin{align}
		\int_{\mathcal{M}}\frac{\dd^2\tau}{\tau_2^2} = \frac{\pi}{3} \, .
	\end{align}
	Then, the analytic part of the one-loop four-graviton amplitude up to eighth order in $\alpha^{\prime 8}$ is
	\begin{align}
		\mathcal{A}_{\text{an}}(s_{ij})&=\frac{2\pi^2}{3}\bigg(1+\frac{1}{3}\zeta_{3}\sigma_3+\frac{29}{180}\zeta_5\sigma_2\sigma_3+\frac{1}{18}\zeta_{3}^2\sigma_3^2-\frac{163}{6\times 7!}\zeta_7\sigma_2^2\sigma_3\nonumber\\&\quad\quad\quad\quad+\frac{79673}{97\times 3^3 \times 6!}\zeta_3\zeta_5\sigma_2\sigma_3^2+\frac{617}{97\times 6^2\times 4!}\zeta_3\zeta_5\sigma_2^4+\mathcal{O}(s_{ij}^9)\bigg)\, .
		\label{eq: analytic}
	\end{align}
	The first line of this equation was already calculated in \cite{DHoker:2019blr,Claasen:2024ssh}.

	\acknowledgments{We thank A. Kleinschmidt, O. Schlotterer, F. Zerbini, D. Dorigoni and A. Edison for useful discussions. We thank A. Kleinschmidt and O. Schlotterer for their valuable comments on the manuscript. MD was supported by the ERE grant RF$\backslash$ERE$\backslash$221103 associated with the Royal Society University Research  Fellowship Grant URF$\backslash$R1$\backslash$221236.}
	
	\appendix
	\section{Laurent polynomial of equivariant iterated Eisenstein integrals}
	\label{appendix: LP}
	The equivariant iterated Eisenstein integrals can be written as
	\begin{align}
		\betaeqvtau{j_1&\ldots&j_\ell\\k_1&\ldots&k_\ell} &=\beta_\Delta^{\rm sv}\big[\begin{smallmatrix}j_1&\ldots&j_\ell\\k_1&\ldots&k_\ell\end{smallmatrix};\tau\big] + \sum_{i=0}^{\ell} \ddsvpure\big[\begin{smallmatrix}j_1&\ldots&j_i\\k_1&\ldots&k_i\end{smallmatrix};\tau\big]
		\beta^{\rm sv}\big[\begin{smallmatrix}j_{i+1}&\ldots&j_\ell\\k_{i+1}&\ldots&k_\ell\end{smallmatrix};\tau\big] \, ,
		\label{beqvTobsv}
	\end{align}
	where the $\beta^{\mathrm{sv}}$ and $\beta^{\mathrm{sv}}_{\Delta}$ were defined in \cite{Gerken:2020yii,Dorigoni:2022npe}. The cuspidal iterated integrals $\beta^{\mathrm{sv}}_\Delta$ do not contribute to the Laurent polynomial. The $d^{\mathrm{sv}}$ are given by \cite{Dorigoni:2022npe,Dorigoni:2024oft}
	\begin{align}
		\label{MGFtoBR32b}
		\ddsv{j_1 &\ldots&j_\ell}{k_1 &\ldots&k_\ell}{\tau} &= \sum_{p_1=0}^{k_1-2-j_1}\cdots\sum_{p_\ell=0}^{k_\ell-2-j_\ell} \frac{ 
			{ k_1{-}2 {-}j_1 \choose p_1} \cdots { k_\ell{-}2 {-}j_\ell \choose p_\ell}}{(4y)^{p_1+\cdots+p_\ell}}  \ccsv{ j_1{+}p_1 &\ldots& j_\ell{+}p_\ell }{k_1 &\ldots &k_\ell}\, ,
	\end{align}
	where $y=\Im\tau$. The $c^{\mathrm{sv}}$ are $\mathbb{Q}$-linear combinations of multiple modular values \cite{Brown:mmv,Brown:2017qwo,Dorigoni:2024oft,Dorigoni:2024iyt,Dorigoni:2022npe}. At depth one, we have \cite{Brown:2017qwo,Dorigoni:2022npe}
	\begin{align}
		\ccsv{j}{k} = - \frac{2\zeta_{k-1}}{k-1}\delta_{j,k-2} \, .
	\end{align}
	These can be found up to degree 20 of depth 2 and 3 in the ancillary file of \cite{Dorigoni:2024oft}. Furthermore, the Laurent polynomial of the $\beta^{\mathrm{sv}}$ for general depth can be inferred from \cite{Broedel:2018izr,Gerken:2020yii} and is given by
	\begin{align}
		\bsvBR{j_1 &\ldots& j_{\ell} }{k_1 &\ldots& k_{\ell} }{\tau}\bigg|_{\mathrm{LP}}
		&= (-4y)^{j_1+\ldots j_{\ell}+\ell}\sum_{p_1=0}^{k_1-2-j_1}\ldots\sum_{p_{\ell}=0}^{k_{\ell}-2-j_{\ell}}
		\frac{B_{k_1}}{k_1!}\ldots\frac{B_{k_\ell}}{k_{\ell}!}\times  \notag \\ 
		& \qquad\qquad\frac{(-1)^{p_1+\ldots +p_{\ell}}{ k_1{-}2 {-}j_1 \choose p_1}\ldots { k_{\ell}{-}2 {-}j_{\ell} \choose p_{\ell}}}{(j_1+p_1+1)\ldots(j_1+p_1+\ldots +j_{\ell}+p_{\ell}+\ell)}\, ,
	\end{align}
	so that the Laurent polynomials of the equivariant iterated Eisenstein integrals are given by 
	\begin{align}
		\betaeqvtau{j_1&\cdots&j_\ell\\k_1&\cdots&k_\ell}\bigg|_{\mathrm{LP}} &= \sum_{i=0}^{\ell} \ddsvpure\big[\begin{smallmatrix}j_1&\cdots&j_i\\k_1&\cdots&k_i\end{smallmatrix};\tau\big]
		\beta^{\rm sv}\big[\begin{smallmatrix}j_{i+1}&\cdots&j_\ell\\k_{i+1}&\cdots&k_\ell\end{smallmatrix};\tau\big]\bigg|_{\mathrm{LP}} \, .
		\label{betaeqvLP}
	\end{align}

	\section{Calculation of depth 4 \texorpdfstring{$\ccsv{j_1&j_2&j_3&j_4}{4&4&4&4}$}{TEXT}}
	\label{appendix: csv}
	The calculations in this appendix heavily rely on the genus one zeta generators and Tsunogai algebra, for which we refer to \cite{Dorigoni:2024iyt}. In order to calculate the constant of the zero mode of weight 8 modular graph functions, we need the rational coefficient multiplying $\zeta_3\zeta_5$ that appears in $c^{\mathrm{sv}}$ of degree 16 and depth 4. These can in turn be calculated using the generating series of $c^{\rm sv}$ \cite{Dorigoni:2024oft}
	\begin{align}
		\mathbb{C}^{\mathrm{sv}}(\{\ep_k\}) &= 1 
		+  \sum_{k_1=4}^{\infty}\sum_{j_1=0}^{k_1-2}   (-1)^{j_1} \frac{(k_1{-}1)}{j_1!} \ccsv{j_1 }{k_1} \ep_{k_1}^{(j_1)}  
		\notag \\
		&\quad
		+ \sum_{k_1=4}^{\infty}\sum_{j_1=0}^{k_1-2}\sum_{k_2=4}^{\infty}\sum_{j_2=0}^{k_2-2}
		(-1)^{j_1+j_2} \frac{(k_1{-}1)(k_2{-}1)}{j_1!j_2!}
		\ccsv{j_1 &j_2 }{k_1 &k_2} \ep_{k_1}^{(j_1)}\ep_{k_2}^{(j_2)} + \ldots\, ,
		\label{CsvGen}
	\end{align}
	where $\ep_k^{(j)}$'s are non-commuting elements of Tsunogai derivation algebra \cite{Tsunogai,Pollack}. This generating series can be expanded in terms of svMZVs, geometric part of the genus one zeta generators $\sigma_i^{\rm g}$ and arithmetic elements $z_i$
	\cite{Dorigoni:2024oft}
	\begin{align}
		\mathbb C^{\rm sv} (\{\epsilon_k\}) &= 1 
		+ 2 \sum_{i_1 \in 2\mathbb N+1} \zeta_{i_1}  \sigmaT^{\rm g}_{i_1} 
		+ 2 \sum_{i_1,i_2 \in 2\mathbb N+1} \zeta_{i_1} \zeta_{i_2} 
		\Big( \sigmaT^{\rm g}_{i_1} \sigmaT^{\rm g}_{i_2} + [ \sigmaT^{\rm g}_{i_1} , \zetaT_{i_2} ] \Big) +\ldots \, .
	\end{align}
	By restricting the MZVs to $\zeta_3\zeta_5$, we find
	\begin{align}
		\mathbb C^{\rm sv} (\{\epsilon_k\})\bigg|_{\zeta_3\zeta_5} &= \zeta_{3} \zeta_{5} 
		\Big( \sigmaT^{\rm g}_{3} \sigmaT^{\rm g}_{5}+\sigmaT^{\rm g}_{5} \sigmaT^{\rm g}_{3} + [ \sigmaT^{\rm g}_{3} , \zetaT_{5} ] + [ \sigmaT^{\rm g}_{5} , \zetaT_{3} ] \Big) \, .
		\label{CgenZ3Z5}
	\end{align}
	The $\GG_4\GG_4\GG_4\GG_4$ sector is related to the cases with four $\ep_4^{(j)}$, therefore we we only write the terms that contribute to this sector. $[\sigma^{\rm g}_3,z_5]$ does not have any contribution. $\sigma^{\rm g}_3$ and $\sigma^{\rm g}_5$ have the following expansion \cite{Dorigoni:2024oft,Dorigoni:2024iyt}
	\begin{align}
		\label{sigma3}
		\sigma_3^{\rm g} &= - \frac{1}{2} \ep_4^{(2)}  + \frac{1}{480} [\ep_4,\ep_4^{(1)}] + \ldots\, ,
		\\
		\label{sigma5}
		\sigma^{\rm g}_5 &= 
		-\tfrac{5  }{48} [\epsilon_4^{(1)},\epsilon_4^{(2)}]
		+\tfrac{1}{5760} [\epsilon_4,\epsilon_6^{(3)}]
		-\tfrac{1}{5760} [\epsilon_4^{(1)},\epsilon_6^{(2)}] +\tfrac{1}{5760} [\epsilon_4^{(2)},\epsilon_6^{(1)}]
		\notag \\
		&\quad
		+\tfrac{1}{3456} [\epsilon_4,[\epsilon_4,\epsilon_4^{(2)}]]
		+\tfrac{1}{6912} [\epsilon_4^{(1)},[\epsilon_4^{(1)},\epsilon_4]] + \ldots \, .
	\end{align}
	The depth two terms of $[z_3,\ep_6]$ can be found in \cite{hain_matsumoto_2020}.
	\begin{align}
		\label{z3e6}
		[z_3,\ep_6] = -\frac{21}{400}[\ep_4,[\ep_4,\ep_4^{(1)}]] + \ldots \, .
	\end{align}
	Then by Jacobi identity and the fact that $\ep_k^{(j)} = \ad_{\ep_0}^{j} \ep_k$ where $\ad_{\ep_0} = [\ep_0,\phantom{A}]$ and also $[z_i,\ep_0]=0$ we have
	\begin{align}
		\label{z3com}
		[\sigma_5,\ep_3] &= \frac{7}{768000} \Big( 4 [\ep_4, [\ep_4^{(1)}, [\ep_4^{(1)}, \ep_4^{(2)}]]] 
		+ 4 [\ep_4, [\ep_4^{(2)}, [\ep_4, \ep_4^{(2)}]]] 
		- 2 [\ep_4^{(1)}, [\ep_4, [\ep_4^{(1)}, \ep_4^{(2)}]]] \notag \\
		&\quad\qquad\qquad + [\ep_4^{(1)}, [\ep_4^{(1)}, [\ep_4, \ep_4^{(2)}]]]
		\Big)+\ldots \, .
	\end{align}
	By substituting (\ref{sigma3}), (\ref{sigma5}) and (\ref{z3com}) into (\ref{CgenZ3Z5}) we can expand $\mathbb{C}^{\rm sv}$ with $\ep_4^{(j)}$'s
	\begin{align}
		\mathbb C^{\rm sv} (\{\epsilon_k\})\bigg|_{\zeta_3\zeta_5}^{\GG_4\GG_4\GG_4\GG_4} =& -\tfrac{7}{884736000} [\ep_4, [\ep_4^{(1)}, [\ep_4^{(1)}, \ep_4^{(2)}]]]  
		- \tfrac{7}{884736000} [\ep_4, [\ep_4^{(2)}, [\ep_4, \ep_4^{(2)}]]] \notag\\&
		+ \tfrac{7}{1769472000} [\ep_4^{(1)}, [\ep_4, [\ep_4^{(1)}, \ep_4^{(2)}]]]  
		- \tfrac{7}{3538944000} [\ep_4^{(1)}, [\ep_4^{(1)}, [\ep_4, \ep_4^{(2)}]]] \notag\\& 
		- \tfrac{1}{2619} [\ep_4, [\ep_4, \ep_4^{(2)}]] \ep_4  
		- \tfrac{1}{4608} [\ep_4, \ep_4^{(1)}] [\ep_4^{(1)}, \ep_4^{(2)}]  
		- \tfrac{1}{5238} [\ep_4^{(1)}, [\ep_4^{(1)}, \ep_4]] \ep_4^{(2)}  
		\notag\\&
		- \tfrac{1}{4608} [\ep_4^{(1)}, \ep_4^{(2)}] [\ep_4, \ep_4^{(1)}]  
		- \tfrac{1}{2619} \ep_4 [\ep_4, [\ep_4, \ep_4^{(2)}]]  \notag\\&
		- \tfrac{1}{5238} \ep_4^{(2)} [\ep_4^{(1)}, [\ep_4^{(1)}, \ep_4]] \, .
		\label{A12}
	\end{align}
	By equating the right hand side of (\ref{CsvGen}) and (\ref{A12}) one can find the $\zeta_3\zeta_5$ terms of $\ccsv{j_1 &j_2 &j_3 &j_4}{4 &4&4&4}$ for different values of $0\geq j_i \geq 2$. The $c^{\rm sv}$'s that appear in the zero mode of (\ref{sigma2sigma3}) and (\ref{sigma2}) have the following values
	\begin{align}
		\ccsv{1 \,1 \,1 \,1}{4 \,4 \,4 \,4} &= 0\,, &
		\ccsv{1 \,1 \,2 \,0}{4 \,4 \,4 \,4} &= \frac{7}{2^{19} 3^6 5^3} \zeta_3 \zeta_5\,, &
		\ccsv{1 \,2 \,0 \,1}{4 \,4 \,4 \,4} &= -\frac{1}{2^8 3^6} \zeta_3 \zeta_5\,,\nonumber \\
		\ccsv{1 \,2 \,1 \,0}{4 \,4 \,4 \,4} &= \frac{127993}{2^{18} 3^6 5^3} \zeta_3 \zeta_5\,, &
		\ccsv{2 \,0 \,1 \,1}{4 \,4 \,4 \,4} &= -\frac{21845107}{2^{19} 3^6 5^3 97} \zeta_3 \zeta_5\,, &
		\ccsv{2 \,0 \,2 \,0}{4 \,4 \,4 \,4} &= \frac{7}{2^{15} 3^7 5^3} \zeta_3 \zeta_5\,,\nonumber \\
		\ccsv{2 \,1 \,0 \,1}{4 \,4 \,4 \,4} &= \frac{11420369}{2^{18} 3^5 5^3 97} \zeta_3 \zeta_5\,, &
		\ccsv{2 \,1 \,1 \,0}{4 \,4 \,4 \,4} &= -\frac{11669107}{2^{17} 3^6 5^3 97} \zeta_3 \zeta_5\,, &
		\ccsv{2 \,2 \,0 \,0}{4 \,4 \,4 \,4} &= -\frac{7}{2^{16} 3^7 5^3} \zeta_3 \zeta_5 \, .
	\end{align}
	
	\providecommand{\href}[2]{#2}\begingroup\raggedright\endgroup


\begin{thebibliography}{10}
		
		\bibitem{Green:2000}
		M.~B. Green and P.~Vanhove, ``{Low-energy expansion of the one-loop type-II
			superstring amplitude},''
		\href{https://dx.doi.org/10.1103/physrevd.61.104011}{{\em Physical Review D}
			{\bfseries 61} no.~10, (Apr., 2000) }.
		\url{http://dx.doi.org/10.1103/PhysRevD.61.104011}.
		
		\bibitem{Green:2008uj}
		M.~B. Green, J.~G. Russo, and P.~Vanhove, ``{Low energy expansion of the
			four-particle genus-one amplitude in type II superstring theory},''
		\href{https://dx.doi.org/10.1088/1126-6708/2008/02/020}{{\em JHEP} {\bfseries
				02} (2008) 020},
		\href{https://arxiv.org/abs/0801.0322}{{\ttfamily arXiv:0801.0322 [hep-th]}}.
		%%CITATION = ARXIV:0801.0322;%%.
		
		\bibitem{DHoker:2015gmr}
		E.~D'Hoker, M.~B. Green, and P.~Vanhove, ``{On the modular structure of the
			genus-one Type II superstring low energy expansion},''
		\href{https://dx.doi.org/10.1007/JHEP08(2015)041}{{\em JHEP} {\bfseries 08}
			(2015) 041}, \href{https://arxiv.org/abs/1502.06698}{{\ttfamily
				arXiv:1502.06698 [hep-th]}}.
		
		\bibitem{DHoker:2019blr}
		E.~D'Hoker and M.~B. Green, ``{Exploring transcendentality in superstring
			amplitudes},'' \href{https://dx.doi.org/10.1007/JHEP07(2019)149}{{\em JHEP}
			{\bfseries 07} (2019) 149},
		\href{https://arxiv.org/abs/1906.01652}{{\ttfamily arXiv:1906.01652
				[hep-th]}}.
		
		\bibitem{DHoker:2015wxz}
		E.~D'Hoker, M.~B. Green, O.~G\"urdogan, and P.~Vanhove, ``{Modular Graph
			Functions},'' \href{https://dx.doi.org/10.4310/CNTP.2017.v11.n1.a4}{{\em
				Commun. Num. Theor. Phys.} {\bfseries 11} (2017) 165--218},
		\href{https://arxiv.org/abs/1512.06779}{{\ttfamily arXiv:1512.06779
				[hep-th]}}.
		
		\bibitem{DHoker:2016mwo}
		E.~D'Hoker and M.~B. Green, ``Identities between modular graph forms,''
		\href{https://dx.doi.org/10.1016/j.jnt.2017.11.015}{{\em J. Number Theory}
			{\bfseries 189} (2018) 25--80},
		\href{https://arxiv.org/abs/1603.00839}{{\ttfamily arXiv:1603.00839 [hep-th]}}.
		%%CITATION = ARXIV:1603.00839;%%.
		
		\bibitem{DHoker:2015sve}
		E.~D'Hoker, M.~B. Green, and P.~Vanhove, ``{Proof of a modular relation between
			1-, 2- and 3-loop Feynman diagrams on a torus},''
		\href{https://dx.doi.org/10.1016/j.jnt.2017.07.022}{{\em J.\ Number Theory}
			(2018) 381},
		\href{https://arxiv.org/abs/1509.00363}{{\ttfamily arXiv:1509.00363 [hep-th]}}.
		%%CITATION = ARXIV:1509.00363;%%.
		
		\bibitem{DHoker:2016quv}
		E.~D'Hoker and J.~Kaidi, ``{Hierarchy of Modular Graph Identities},''
		\href{https://dx.doi.org/10.1007/JHEP11(2016)051}{{\em JHEP} {\bfseries 11}
			(2016) 051},
		\href{https://arxiv.org/abs/1608.04393}{{\ttfamily arXiv:1608.04393 [hep-th]}}.
		%%CITATION = ARXIV:1608.04393;%%.
		
		\bibitem{Gerken:2018zcy}
		J.~E. Gerken and J.~Kaidi, ``{Holomorphic subgraph reduction of higher-point
			modular graph forms},''
		\href{https://dx.doi.org/10.1007/JHEP01(2019)131}{{\em JHEP} {\bfseries 01}
			(2019) 131},
		\href{https://arxiv.org/abs/1809.05122}{{\ttfamily arXiv:1809.05122 [hep-th]}}.
		%%CITATION = ARXIV:1809.05122;%%.
		
		\bibitem{Basu:2016kli}
		A.~Basu, ``{Proving relations between modular graph functions},''
		\href{https://dx.doi.org/10.1088/0264-9381/33/23/235011}{{\em Class. Quant.
				Grav.} {\bfseries 33} no.~23, (2016) 235011},
		\href{https://arxiv.org/abs/1606.07084}{{\ttfamily arXiv:1606.07084 [hep-th]}}.
		%%CITATION = ARXIV:1606.07084;%%.
		
		\bibitem{Gerken:2020aju}
		J.~E. Gerken, ``{Basis Decompositions and a Mathematica Package for Modular
			Graph Forms},'' \href{https://dx.doi.org/10.1088/1751-8121/abbdf2}{{\em J.
				Phys. A} {\bfseries 54} no.~19, (2021) 195401},
		\href{https://arxiv.org/abs/2007.05476}{{\ttfamily arXiv:2007.05476
				[hep-th]}}.
		
		\bibitem{Kleinschmidt:2017ege}
		A.~Kleinschmidt and V.~Verschinin, ``{Tetrahedral modular graph functions},''
		\href{https://dx.doi.org/10.1007/JHEP09(2017)155}{{\em JHEP} {\bfseries 09}
			(2017) 155}, \href{https://arxiv.org/abs/1706.01889}{{\ttfamily
				arXiv:1706.01889 [hep-th]}}.
		
		\bibitem{Brown:2017qwo}
		F.~Brown, ``{A class of non-holomorphic modular forms I},''
		\href{https://dx.doi.org/10.1007/s40687-018-0130-8}{{\em Res. Math. Sci.}
			{\bfseries 5} (2018) 5:7}, \href{https://arxiv.org/abs/1707.01230}{{\ttfamily
				arXiv:1707.01230 [math.NT]}}.
		
		\bibitem{Brown:2017qwo2}
		F.~Brown, ``{A class of non-holomorphic modular forms II : equivariant iterated
			Eisenstein integrals},'' \href{https://dx.doi.org/10.1017/fms.2020.24}{{\em
				Forum~of~Mathematics,~Sigma} {\bfseries 8} (2020) 1},
		\href{https://arxiv.org/abs/1708.03354}{{\ttfamily arXiv:1708.03354
				[math.NT]}}.
		%%CITATION = ARXIV:1707.01230;%%.
		
		\bibitem{Broedel:2018izr}
		J.~Broedel, O.~Schlotterer, and F.~Zerbini, ``{From elliptic multiple zeta
			values to modular graph functions: open and closed strings at one loop},''
		\href{https://dx.doi.org/10.1007/JHEP01(2019)155}{{\em JHEP} {\bfseries 01}
			(2019) 155}, \href{https://arxiv.org/abs/1803.00527}{{\ttfamily
				arXiv:1803.00527 [hep-th]}}.
		
		\bibitem{Gerken:2019cxz}
		J.~E. Gerken, A.~Kleinschmidt, and O.~Schlotterer, ``{All-order differential
			equations for one-loop closed-string integrals and modular graph forms},''
		\href{https://dx.doi.org/10.1007/JHEP01(2020)064}{{\em JHEP} {\bfseries 01}
			(2020) 064},
		\href{https://arxiv.org/abs/1911.03476}{{\ttfamily arXiv:1911.03476 [hep-th]}}.
		%%CITATION = ARXIV:1911.03476;%%.
		
		\bibitem{Gerken:2020yii}
		J.~E. Gerken, A.~Kleinschmidt, and O.~Schlotterer, ``{Generating series of all
			modular graph forms from iterated Eisenstein integrals},''
		\href{https://dx.doi.org/10.1007/JHEP07(2020)190}{{\em JHEP} {\bfseries 07}
			no.~07, (2020) 190}, \href{https://arxiv.org/abs/2004.05156}{{\ttfamily
				arXiv:2004.05156 [hep-th]}}.
		
		\bibitem{drewitt2022laplace}
		J.~Drewitt, ``Laplace-eigenvalue equations for length three modular iterated
		integrals,'' \href{https://dx.doi.org/10.1016/j.jnt.2021.11.005}{{\em Journal
				of Number Theory} {\bfseries 239} (2022) 78--112},
		\href{https://arxiv.org/abs/2104.09916}{{\ttfamily arXiv:2104.09916 [math]}}.
		
		\bibitem{Dorigoni:2022npe}
		D.~Dorigoni, M.~Doroudiani, J.~Drewitt, M.~Hidding, A.~Kleinschmidt,
		N.~Matthes, O.~Schlotterer, and B.~Verbeek, ``{Modular graph forms from
			equivariant iterated Eisenstein integrals},''
		\href{https://dx.doi.org/10.1007/JHEP12(2022)162}{{\em JHEP} {\bfseries 12}
			(2022) 162}, \href{https://arxiv.org/abs/2209.06772}{{\ttfamily
				arXiv:2209.06772 [hep-th]}}.
		
		\bibitem{Dorigoni:2024oft}
		D.~Dorigoni, M.~Doroudiani, J.~Drewitt, M.~Hidding, A.~Kleinschmidt,
		O.~Schlotterer, L.~Schneps, and B.~Verbeek, ``{Non-holomorphic modular forms
			from zeta generators},''
		\href{https://dx.doi.org/10.1007/JHEP10(2024)053}{{\em JHEP} {\bfseries 10}
			(2024) 53}, \href{https://arxiv.org/abs/2403.14816}{{\ttfamily
				arXiv:2403.14816 [hep-th]}}.
		
		\bibitem{Gerken:2020xte}
		J.~E. Gerken, \href{https://dx.doi.org/10.18452/21829}{{\em {Modular Graph
					Forms and Scattering Amplitudes in String Theory}}}.
		\newblock PhD thesis, Humboldt U., Berlin, Humboldt U., Berlin, 2020.
		\newblock \href{https://arxiv.org/abs/2011.08647}{{\ttfamily arXiv:2011.08647
				[hep-th]}}.
		
		\bibitem{Hidding:2022vjf}
		M.~Hidding, O.~Schlotterer, and B.~Verbeek, ``{Elliptic modular graph forms II:
			Iterated integrals},'' \href{https://arxiv.org/abs/2208.11116}{{\ttfamily
				arXiv:2208.11116 [hep-th]}}.
		
		\bibitem{Goncharov:2010jf}
		A.~B. Goncharov, M.~Spradlin, C.~Vergu, and A.~Volovich, ``{Classical
			Polylogarithms for Amplitudes and Wilson Loops},''
		\href{https://dx.doi.org/10.1103/PhysRevLett.105.151605}{{\em Phys. Rev.
				Lett.} {\bfseries 105} (2010) 151605},
		\href{https://arxiv.org/abs/1006.5703}{{\ttfamily arXiv:1006.5703 [hep-th]}}.
		%%CITATION = ARXIV:1006.5703;%%.
		
		\bibitem{Zerbini:2015rss}
		F.~Zerbini, ``{Single-valued multiple zeta values in genus 1 superstring
			amplitudes},'' \href{https://dx.doi.org/10.4310/CNTP.2016.v10.n4.a2}{{\em
				Commun. Num. Theor. Phys.} {\bfseries 10} (2016) 703--737},
		\href{https://arxiv.org/abs/1512.05689}{{\ttfamily arXiv:1512.05689
				[hep-th]}}.
		
		\bibitem{DHoker:2019xef}
		E.~D'Hoker and M.~B. Green, ``{Absence of irreducible multiple zeta-values in
			melon modular graph functions},''
		\href{https://dx.doi.org/10.4310/CNTP.2020.v14.n2.a2}{{\em Commun. Num.
				Theor. Phys.} {\bfseries 14} no.~2, (2020) 315--324},
		\href{https://arxiv.org/abs/1904.06603}{{\ttfamily arXiv:1904.06603 [hep-th]}}.
		%%CITATION = ARXIV:1904.06603;%%.
		
		\bibitem{Zagier:2019eus}
		D.~Zagier and F.~Zerbini, ``{Genus-zero and genus-one string amplitudes and
			special multiple zeta values},''
		\href{https://dx.doi.org/10.4310/CNTP.2020.v14.n2.a4}{{\em Commun. Num.
				Theor. Phys.} {\bfseries 14} no.~2, (2020) 413--452},
		\href{https://arxiv.org/abs/1906.12339}{{\ttfamily arXiv:1906.12339
				[math.NT]}}.
		%%CITATION = ARXIV:1906.12339;%%.
		
		\bibitem{Vanhove:2020qtt}
		P.~Vanhove and F.~Zerbini, ``{Building blocks of closed and open string
			amplitudes},'' \href{https://dx.doi.org/10.22323/1.383.0022}{{\em PoS}
			{\bfseries MA2019} (2022) 022},
		\href{https://arxiv.org/abs/2007.08981}{{\ttfamily arXiv:2007.08981
				[hep-th]}}.
		
		\bibitem{Dorigoni:2021jfr}
		D.~Dorigoni, A.~Kleinschmidt, and O.~Schlotterer, ``{Poincar\'e series for
			modular graph forms at depth two. Part I. Seeds and Laplace systems},''
		\href{https://dx.doi.org/10.1007/JHEP01(2022)133}{{\em JHEP} {\bfseries 01}
			(2022) 133}, \href{https://arxiv.org/abs/2109.05017}{{\ttfamily
				arXiv:2109.05017 [hep-th]}}.
		
		\bibitem{Dorigoni:2021ngn}
		D.~Dorigoni, A.~Kleinschmidt, and O.~Schlotterer, ``{Poincar\'e series for
			modular graph forms at depth two. Part II. Iterated integrals of cusp
			forms},'' \href{https://dx.doi.org/10.1007/JHEP01(2022)134}{{\em JHEP}
			{\bfseries 01} (2022) 134},
		\href{https://arxiv.org/abs/2109.05018}{{\ttfamily arXiv:2109.05018
				[hep-th]}}.
		
		\bibitem{Dorigoni:2024iyt}
		D.~Dorigoni, M.~Doroudiani, J.~Drewitt, M.~Hidding, A.~Kleinschmidt,
		O.~Schlotterer, L.~Schneps, and B.~Verbeek, ``{Canonicalizing zeta
			generators: genus zero and genus one},''
		\href{https://arxiv.org/abs/2406.05099}{{\ttfamily arXiv:2406.05099
				[math.QA]}}.
		
		\bibitem{Claasen:2024ssh}
		E.~Claasen and M.~Doroudiani, ``{Transcendentality of Type II superstring
			amplitude at one-loop},'' \href{https://arxiv.org/abs/2412.04381}{{\ttfamily
				arXiv:2412.04381 [hep-th]}}.
		
		\bibitem{Edison:2021ebi}
		A.~Edison, M.~Guillen, H.~Johansson, O.~Schlotterer, and F.~Teng, ``{One-loop
			matrix elements of effective superstring interactions:
			\ensuremath{\alpha}'-expanding loop integrands},''
		\href{https://dx.doi.org/10.1007/JHEP12(2021)007}{{\em JHEP} {\bfseries 12}
			(2021) 007}, \href{https://arxiv.org/abs/2107.08009}{{\ttfamily
				arXiv:2107.08009 [hep-th]}}.
		
		\bibitem{DHoker:2022dxx}
		E.~D'Hoker and J.~Kaidi, ``{Lectures on modular forms and strings},''
		\href{https://arxiv.org/abs/2208.07242}{{\ttfamily arXiv:2208.07242
				[hep-th]}}.
		
		\bibitem{serre2012course}
		J.-P. Serre, {\em A course in arithmetic}, vol.~7.
		\newblock Springer Science \& Business Media, 2012.
		
		\bibitem{DHoker:2017zhq}
		E.~D'Hoker and W.~Duke, ``Fourier series of modular graph functions,''
		\href{https://dx.doi.org/10.1016/j.jnt.2018.04.012}{{\em J. Number Theory}
			{\bfseries 192} (2018) 1--36},
		\href{https://arxiv.org/abs/1708.07998}{{\ttfamily arXiv:1708.07998
				[math.NT]}}.
		
		\bibitem{Maass}
		H.~Maass, {\em Lectures on modular functions of one complex variable}, vol.~29
		of {\em Tata Institute of Fundamental Research Lectures on Mathematics and
			Physics}.
		\newblock Tata Institute of Fundamental Research, Bombay, second~ed., 1983.
		\newblock With notes by Sunder Lal.
		
		\bibitem{Zagierunpub}
		D.~Zagier, ``Unpublished; notes on lattice sums.''
		\newblock Unpusblished.
		
		\bibitem{Fay}
		J.~D. Fay, {\em Theta functions on {R}iemann surfaces}.
		\newblock Lecture Notes in Mathematics, Vol. 352. Springer-Verlag, Berlin-New
		York, 1973.
		
		\bibitem{Brown:2014pnb}
		F.~Brown, ``{Multiple Modular Values and the relative completion of the
			fundamental group of $M_{1,1}$},''
		\href{https://arxiv.org/abs/1407.5167}{{\ttfamily arXiv:1407.5167
				[math.NT]}}.
		
		\bibitem{brown2017a}
		F.~Brown, ``A class of non-holomorphic modular forms {{III}}: Real analytic
		cusp forms for {$\mathrm{SL}_2(\mathbb{Z})$},''
		\href{https://dx.doi.org/10.1007/s40687-018-0151-3}{{\em Research in the
				Mathematical Sciences} {\bfseries 5} no.~3, (2018) Paper No. 34, 36},
		\href{https://arxiv.org/abs/1710.07912}{{\ttfamily arXiv:1710.07912}}.
		
		\bibitem{Pollack}
		A.~Pollack, ``{Relations between derivations arising from modular forms},''
		\url{https://dukespace.lib.duke.edu/dspace/handle/10161/1281}, 2009.
		\newblock Undergraduate thesis, Duke University.
		
		\bibitem{Tsunogai}
		H.~Tsunogai, ``On some derivations of {L}ie algebras related to {G}alois
		representations,'' \href{https://dx.doi.org/10.2977/prims/1195164794}{{\em
				Publ. Res. Inst. Math. Sci.} {\bfseries 31} no.~1, (1995) 113--134}.
		
		\bibitem{matthes2017algebraic}
		N.~Matthes, ``On the algebraic structure of iterated integrals of quasimodular
		forms,'' \href{https://dx.doi.org/10.2140/ant.2017.11.2113}{{\em Algebra \&
				Number Theory} {\bfseries 11} no.~9, (2017) 2113--2130},
		\href{https://arxiv.org/abs/math-NT/1708.04561}{{\ttfamily
				arXiv:math-NT/1708.04561}}.
		
		\bibitem{Green:1999pv}
		M.~B. Green and P.~Vanhove, ``{The Low-energy expansion of the one loop type II
			superstring amplitude},''
		\href{https://dx.doi.org/10.1103/PhysRevD.61.104011}{{\em Phys.Rev.}
			{\bfseries D61} (2000) 104011},
		\href{https://arxiv.org/abs/hep-th/9910056}{{\ttfamily arXiv:hep-th/9910056
				[hep-th]}}.
		%%CITATION = HEP-TH/9910056;%%.
		
		\bibitem{Green:1981yb}
		M.~B. Green and J.~H. Schwarz, ``{Supersymmetrical String Theories},''
		\href{https://dx.doi.org/10.1016/0370-2693(82)91110-8}{{\em Phys. Lett. B}
			{\bfseries 109} (1982) 444--448}.
		
		\bibitem{Brown:mmv}
		F.~Brown, ``{Multiple modular values and the relative completion of the
			fundamental group of ${\cal M}_{1,1}$},''
		\href{https://arxiv.org/abs/1407.5167}{{\ttfamily arXiv:1407.5167
				[math.NT]}}.
		
		\bibitem{hain_matsumoto_2020}
		R.~Hain and M.~Matsumoto, ``Universal mixed elliptic motives,''
		\href{https://dx.doi.org/10.1017/S1474748018000130}{{\em Journal of the
				Institute of Mathematics of Jussieu} {\bfseries 19} no.~3, (2020) 663--766},
		\href{https://arxiv.org/abs/1512.03975}{{\ttfamily arXiv:1512.03975
				[math.AG]}}.
		
	\end{thebibliography}
\end{document}